\documentclass[aps,prb,reprint,superscriptaddress,showpacs,floatfix,amsmath,amssymb]{revtex4-1}
\usepackage{graphicx}

\DeclareMathOperator{\trace}{Tr}
\newcommand{\nn}[0]{ \nonumber \\ }
\renewcommand{\vec}[1]{ \boldsymbol{#1} }
\renewcommand{\vr}[0]{ \vec{r} }
\newcommand{\IM}[0]{ \mathrm{Im} }
\newcommand{\RE}[0]{ \mathrm{Re} }
\newcommand{\Res}[0]{ \mathrm{Res} }

\newcommand{\indll}[3]{ \int_{#2}^{#3}\!\!\!\!\mathrm{d}#1 \; }
\newcommand{\cH}[0]{ \hat{ \mathcal{H} } }
\newcommand{\cD}[0]{ \hat{ \mathcal{D} } }
\newcommand{\cN}[0]{ \hat{ \mathcal{N} } }
\newcommand{\FF}[1]{ \hat{ \phi }_{#1}^{\phantom{\dagger}} }
\newcommand{\FFd}[1]{ \hat{ \phi }_{#1}^{ \dagger } }
\newcommand{\FFi}[1]{ \hat{ d }_{#1}^{\phantom{\dagger}} }
\newcommand{\FFid}[1]{ \hat{ d }_{#1}^{ \dagger } }
\newcommand{\av}[1]{ \left\langle #1 \right\rangle }
\newcommand{\probe}[0]{ \mathrm{probe} }

\begin{document}

\title{Luttinger-field approach to thermoelectric transport in nanoscale conductors}

\author{F. G. Eich}
\email[]{eichf@missouri.edu}
\affiliation{Department of Physics, University of Missouri-Columbia, Columbia, Missouri 65211, USA}

\author{A. Principi}
\affiliation{Department of Physics, University of Missouri-Columbia, Columbia, Missouri 65211, USA}

\author{M. \surname{Di Ventra}}
\affiliation{Department of Physics, University of California, San Diego, La Jolla, California 92093, USA}

\author{G. Vignale}
\affiliation{Department of Physics, University of Missouri-Columbia, Columbia, Missouri 65211, USA}

\date{\today}

\begin{abstract}
  Thermoelectric transport in nanoscale conductors is analyzed in terms of the response of the system to a thermo-mechanical field,
  first introduced by Luttinger, which couples to the electronic energy density. While in this approach the temperature remains spatially
  uniform, we show that a spatially varying thermo-mechanical field effectively simulates a temperature gradient across the system and
  allows us to calculate the electric and thermal currents that flow due to the thermo-mechanical field. In particular, we show that, in the
  long-time limit, the currents thus calculated reduce to those that one obtains from the Landauer-B{\"u}ttiker formula, suitably
  generalized to allow for different temperatures in the reservoirs, if the thermo-mechanical field is applied to prepare the system, and subsequently turned off at ${t=0}$.
  Alternatively, we can drive the system out of equilibrium by switching the thermo-mechanical field after the initial
  preparation. We compare these two scenarios, employing a model noninteracting Hamiltonian, in the linear regime, in which they coincide, and in the
  nonlinear regime in which they show marked differences. We also show how an operationally defined local effective temperature can be computed within this formalism.
\end{abstract}


\maketitle

\section{ Introduction } \label{SEC:introduction}

The problem of calculating the thermal and electrical transport properties of nanoscale conductors has recently attracted great interest
in the context of growing efforts to achieve efficient conversion of heat into electricity, and vice versa.\cite{NolasGoldsmid:01,DubiDiVentra:11}
On the theoretical side, the field is riddled with  conceptual difficulties that can be traced back to the very foundations of statistical physics. Concepts like
temperature, heat current, and thermal conductivity were originally defined at the macroscopic level, or in quasi-equilibrium situations
in which they vary slowly in space and time. How are we to define these concepts at the nanoscale, where the laws of quantum mechanics
take hold, and where the above-mentioned quantities are likely to exhibit rapid variations, both in space and in time? One of the main
theoretical questions is how to convert the temperature, originally defined as a statistical parameter governing the equilibrium of energy
exchanges between different parts of a macroscopic system, into a dynamical field coupling to mechanical degrees of freedom, which can
be driven strongly out of equilibrium. The recent development of scanning thermal microscopy,\cite{Majumdar:99,YuKim:11,KimLee:11,KimReddy:12,MengesGotsmann:12} 
allowing for measurements of a local effective temperature on the atomic scale, provides additional strong motivation for seeking a sharp answer to the above questions.

Many years ago, Luttinger took a first step in this direction by proposing that the thermoelectric transport properties of a
macroscopic electron liquid could be calculated by subjecting the system to a space- and time-varying field $\psi(\vr,t)$.\cite{Luttinger:64a}
The $\psi$ field was to be linearly coupled to the energy density, for which Luttinger chose one of several possible definitions--all
equivalent in the long-wavelength limit. Luttinger's idea was that the dynamical response of the system to the varying field $\psi$ would
be  equivalent to the response to a temperature gradient in situations in which the latter is slowly varying, but would extend the concept
of thermal response to situations in which the traditional notion of temperature is no longer meaningful. Noting the similarity to Einstein's
theory of gravity--a field coupling to the energy density--Luttinger dubbed his $\psi$ field a ``gravitational field''--in a purely formal sense
of course.  We prefer to call it ``thermo-mechanical'' (TM) field, since it acts, in a very precise sense, as the mechanical proxy for the
temperature. The gradient of this field drives the thermal current, just as the gradient of the electric potential drives the electric current.

In the half century elapsed since the publication of the original paper, Luttinger's idea has found several applications in the calculation of
the linear response of macroscopic systems.\cite{Shastry:09} In a recent paper, we have shown that the TM field offers a natural path to the
inclusion of thermoelectric effects in a general-purpose time-dependent density-functional theory.\cite{EichVignale:14} However, to date there are no reported
applications of these ideas to nanoscale conductors--and this in spite of very significant progress in the theoretical treatment of these systems.

One of the most successful models of transport at the nanoscale is the Landauer-B\"uttiker formalism (LB),\cite{Landauer:57,BuettikerPinhas:85,Landauer:89} in which the nanoscale system is
assumed to be connected, via ideal leads, to several reservoirs independently in equilibrium at different chemical potentials $\mu_{\alpha}$ and
temperatures $T_{\alpha}$ ($\alpha=1,\ldots,N$, where $N$ is the number of reservoirs). The electric and thermal currents, in the LB approach, are expressed
in terms of the quantum mechanical transmission probabilities from each  lead into the others, and the equilibrium distribution functions
of the reservoirs.\cite{DiVentra:08} Clearly, there is no room for any dynamical behavior of the temperature in this approach. In fact, there is
no room for any dynamical effects at all, since the transmission probabilities are calculated from an effective mean field that does not fluctuate
in time, thus ruling out inelastic many-body effects.\cite{SaiDiVentra:05,KoentoppEvers:06,VignaleDiVentra:09}
Note that we employ here a strict definition of dynamical effects, i.e., effects on time scales shorter than the typical equilibration time. It is possible to employ the
LB formalism to address time-dependent thermoelectric transport in the opposite regime when the system evolves adiabatically.\cite{DiVentra:08,SanchezLopez:13,LimSanchez:13}

Efforts to go beyond the LB formulation typically involve the use of nonequilibrium many-body theory (Keldysh formalism).\cite{Keldysh:65,DiVentra:08,StefanucciVanLeeuwen:13} 
An outstanding result obtained through this formalism is the Meir-Wingreen formula, \cite{MeirWingreen:92,WingreenMeir:93,JauhoMeir:94}
expressing the electric current in terms of the exact interacting Green's function
for the nanoscale system and self-energies arising from its coupling to the leads. At variance with the LB approach, the full dynamics--from the initial
preparation to the tentative steady state--of the nanoscale system is described. To this end, the device is assumed to
be initially decoupled from the leads, which are in equilibrium with reservoirs at different chemical potentials $\mu_{\alpha}$ and temperatures $T_{\alpha}$. At time $t=0$,
the coupling between the system and the reservoir is established and the long-time behavior of the currents
is calculated. Clearly this approach, while fully dynamical in the treatment of the currents, continues to treat the temperature as a static
thermodynamic variable, which controls the population of the electronic states in the reservoirs. In the following, we will refer to this approach to the
transport problem as the LB approach since it can be shown that it coincides with the purely static LB approach if a steady state is reached.

In this paper, we present the first application of Luttinger's $\psi$ field idea to the calculation of thermal transport through a nanoscale junction.
The basic idea can be illustrated by considering the occupation functions in the LB approach $f_\alpha = f( \epsilon/ k_{\mathrm{B}} T_\alpha )$.
They are different in the leads, labeled by $\alpha$, since the temperatures are chosen differently. In Luttinger's approach the difference in the occupation
functions is achieved by rescaling the energy, i.e., $f_\alpha = f( \lambda_\alpha \epsilon / k_{\mathrm{B}} T)$. The relation of the scaling factor $\lambda_\alpha = T / T_\alpha$ to the
TM fields in the leads depends on whether the TM fields are applied during the initial preparation, which means that the system is allowed to reach
equilibrium in the presence of the TM fields, or switched on at the beginning of the time evolution.
This means that we completely replace the different temperatures in the reservoirs by TM fields. The conventional statistical temperature
remains constant throughout the system. To calculate the currents, we closely follow the formulation of the nonequilibrium Green's function theory
introduced by Cini\cite{Cini:80} and developed by Stefanucci and Almbladh.\cite{StefanucciAlmbladh:04} In this approach the coupling between the system and the leads exists from $t=-\infty$,
and--in this sense--the system is said to be ``partition free.'' The leads and the nanoscale system are initially in equilibrium with a unique reservoir at a
chemical potential $\mu$ and temperature $T$. At time $t=0$, different electric potentials $U_{\alpha}$ and TM fields $\psi_{\alpha}$ are
applied to the leads. We show that this leads to the identification $\lambda_\alpha = (1 + \psi_\alpha)^{-1}$ or equivalently $T_\alpha = (1 + \psi_\alpha) T$.
The resulting electric and thermal currents are calculated in the long-time limit. Our main result is that, for a noninteracting system, in the linear response
regime, the current calculated in this manner coincides with the current calculated in the LB approach.
Furthermore, we demonstrate that the LB result can be fully recovered in the nonlinear regime, if the TM fields are
applied during the initial preparation of the system, and turned off at ${t=0}$. This is certainly good news, which builds confidence in the general applicability of
Luttinger's approach to nanoscale conductors. In this case, we find $\lambda_\alpha = (1 + \psi_\alpha)$, which implies
$T_\alpha = T / (1 + \psi_\alpha)$.

Many-body effects are not included here, but we expect to be able to handle them, at least approximately, through the recently introduced formalism
of thermal density-functional theory.\cite{EichVignale:14} The dynamical (retarded) nature of the effective potentials is expected to introduce dynamical corrections quite
analogous to the ones discussed in Refs.\ \onlinecite{VignaleDiVentra:09,KurthStefanucci:13} for charge transport.
Furthermore, we discuss a common procedure to define an effective local temperature for nanoscale systems: a local temperature as obtained by computing the TM field that
must be applied to a thermal probe lead, in order to suppress the flow of thermal current between the probe and the system. We leave the comparison of this local temperature
with other alternative definitions\cite{DubiDiVentra:11} for future work.

This paper is organized as follows. In Sec.\ \ref{SEC:ThermoelectricTransport}, we introduce the model Hamiltonian employed to obtain the
formal expressions for the currents in the partition-free scheme. In Sec.\ \ref{SEC:SteadyState}, we compute the long-time limit of
the currents and show that it agrees with the results of the LB formalism in the linear response regime. Details of the calculations are
presented in Appendices \ref{APP:particleCurrent}, \ref{APP:memoryLoss}, and \ref{APP:energyCurrent}. In Sec.\ \ref{SEC:LBvsTM}, we compare the
LB to the TM approach to thermal transport highlighting the differences, which appear when one goes beyond the linear response approximation.
In Sec.\ \ref{SEC:localTemperature}, an operational definition of the local effective temperature--by virtue of a local probe--is
calculated within the TM field formalism. In Sec.\ \ref{SEC:conclusion}, we summarize our findings
and briefly discuss how the effect of interactions can be included within the framework of thermal density-functional theory.

\section{ Thermoelectric transport in nanoscale junctions } \label{SEC:ThermoelectricTransport}

\begin{figure}
  \includegraphics[width=.48\textwidth]{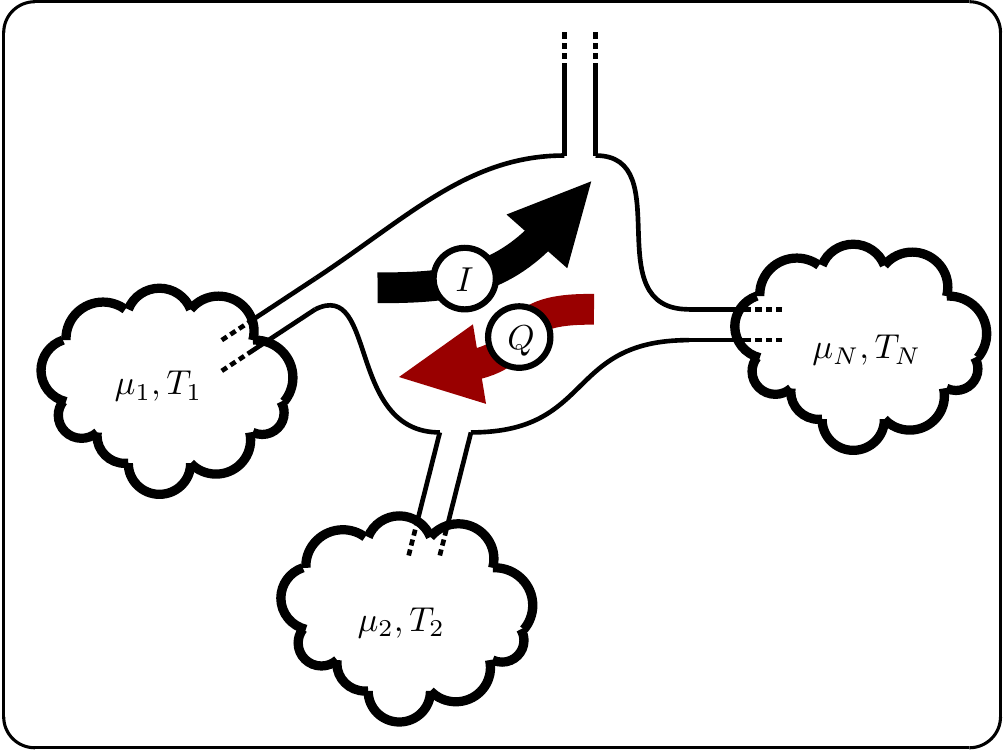}
  \caption{(Color online) This sketch shows a typical transport setup where a nanoscale junction (central region) is
    connected via leads to reservoirs. If the leads are held at different potentials ${\mu_{1}, \mu_{2}, \ldots, \mu_{N}}$
    and/or different temperatures ${T_{1}, T_{2}, \ldots, T_{N}}$, a charge current ${I}$ and a heat current ${Q}$
    will flow through the junction. \label{FIG:LeadsJunction}}
\end{figure}
In order to investigate the thermal and electric transport through a nanoscale junction, we consider a junction coupled to
reservoirs via conducting leads. This setup is shown in Fig.\ \ref{FIG:LeadsJunction}. The conducting leads are modeled by  mean-field Hamiltonians
\begin{align}
  \cH_{\alpha} = \sum_k \epsilon_{\alpha, k} \FFd{\alpha, k} \FF{\alpha, k} ~, \label{leadHamiltonian}
\end{align}
where ${\alpha}$ labels the leads connected to the nanoscale junction. In specific calculations, we will model the leads by an
infinite tight-binding chain,
\begin{subequations} \label{leadsChain}
  \begin{align}
    & \cH_{\alpha} = t_{\alpha} \sum_{i=0}^{N-2} \left( \FFd{\alpha, i+1} \FF{\alpha, i} + \FFd{\alpha, i} \FF{\alpha, i+1} \right)  \label{H_leads_chain_discrete} \\
    & \xrightarrow[N \to \infty]{} \tfrac{2}{\pi} \indll{q}{0}{\pi} 2 t_{\alpha} \cos(q) \FFd{\alpha, q} \FF{\alpha, q} ~, \label{H_leads_chain_limit}
  \end{align}
\end{subequations}
i.e., the leads are characterized by a single band with bandwidth ${4 t_{\alpha}}$. The junction is governed by a Hamiltonian
\begin{align}
  \cH_{\mathrm{imp}} & = \sum_n \epsilon_{n} \FFid{n} \FFi{n} ~, \label{impurityHamiltonian}
\end{align}
where $n$ labels the eigenstates of the microscopic device connected to the leads, e.g., the energy levels of a molecule.
For explicit calculations, we consider a single state $\FFi{}$ in the junction, i.e., a single impurity level at energy $\epsilon_{0}$.
The contact between the junction and the leads is modeled by tunneling amplitudes $V_{(\alpha, k), n}$ between state $k$ in lead ${\alpha}$
and the energy level $n$ in the junction. This contribution to the Hamiltonian reads,
\begin{align}
  \cH_{\alpha, \mathrm{imp}} & = \sum_{k, n} \left( V_{(\alpha, k), n} \FFd{\alpha, k} \FFi{n} + V^\star_{(\alpha, k), n} \FFid{n} \FF{\alpha, k} \right) ~. \label{couplingHamiltonian}
\end{align}
Again, for explicit calculations we consider a single impurity site that only couples to the closest site (taken to be $i = 0$) of the tight-binding chain, i.e.,
\begin{subequations} \label{leadsChainImpurtiyCoupling}
  \begin{align}
    & \cH_{\alpha, \mathrm{imp}} = V_{\alpha} \FFd{\alpha 0} \FFi{} + V^\star_{\alpha} \FFid{} \FF{\alpha 0} \label{H_leads_chain_impurity_coupling_discrete} \\
    & \xrightarrow[N \to \infty]{} \tfrac{2}{\pi} \indll{q}{0}{\pi} \sin(q) \left( V_{\alpha} \FFd{\alpha, q} \FFi{} + V^\star_{\alpha} \FFid{} \FF{\alpha, q} \right)
    ~. \label{H_leads_chain_impurity_coupling_limit}
  \end{align}
\end{subequations}

As already mentioned in the introduction we will follow the partition-free approach to transport in order to calculate the charge and heat current
induced by a bias in the electric \emph{and} the TM potential. This means that the initial state of our system is given by the equilibrium statistical operator
(density matrix),
\begin{align}
  \cD & = \frac{\exp^{ - \beta \left(\cH_{0} - \mu \cN \right) }}{\mathcal{Z}} \;\; , \;\; \mathcal{Z} = \trace\left[\exp^{ - \beta \left( \cH_{0} - \mu \cN \right)}\right] ~, \label{equilibriumDM}
\end{align}
defined with respect to the Hamiltonian
\begin{align}
  \cH_{0} & = \sum_{\alpha} \left( \cH_{\alpha} + \cH_{\alpha, \mathrm{imp}} \right) + \cH_{\mathrm{imp}} ~. \label{equilibriumHamiltonian}
\end{align}
Note that $\beta = \frac{1}{k_{\mathrm{B}}T}$ and $\mu$ are the inverse temperature and the chemical potential of the initial equilibrium, i.e., they are identical for the
entire system composed of the leads and the junction. At the initial time $t = t_{0}$, we switch external electric and TM fields in the leads, i.e., we perturb the equilibrium
state at ${t = t_{0}}$ by
\begin{align}
  \cH_{1} & = \sum_{\alpha, k} \big( U_{\alpha} + \psi_{\alpha} \left(\epsilon_{\alpha, k} - \mu + U_{\alpha}\right) \big) \FFd{\alpha, k} \FF{\alpha, k} ~, \label{biasHamiltonian}
\end{align}
where $U_{\alpha}$ and ${\psi_{\alpha}}$ are the electric and the TM fields in the leads, respectively. This means that while in the initial equilibrium
the dispersion in the leads is given by
\begin{align}
  \bar\epsilon_{\alpha, k} = \epsilon_{\alpha, k} - \mu ~, \label{equilibriumDispersion}
\end{align}
the dispersion in the leads during the time propagation is given by
\begin{align}
  \tilde\epsilon_{\alpha, k} = \left(1 + \psi_{\alpha}\right) \left( \epsilon_{\alpha, k} - \mu + U_{\alpha} \right) ~. \label{propagationDispersion}
\end{align}

The particle current $I_{\alpha}$ and the energy current $J_{\alpha}$ in lead ${\alpha}$ are defined via the time derivative
of the particle-number density and the energy density in the lead, respectively,
\begin{subequations} \label{currentsDefinition}
  \begin{align}
    I_{\alpha}(t) & = - \partial_t n_{\alpha}(t) = - \partial_t \sum_{k} \av{ \FFd{\alpha, k}(t) \FF{\alpha, k}(t) } ~, \label{I} \\
    J_{\alpha}(t) & = - \partial_t h_{\alpha}(t) = - \partial_t \sum_{k} \av{ \bar \epsilon_{\alpha, k} \FFd{\alpha, k}(t) \FF{\alpha, k}(t) } ~. \label{J}
  \end{align}
\end{subequations}
The expectation value is taken with respect to $\cD$ given by Eq.\ \eqref{equilibriumDM}, hence $\av{\ldots} = \trace\left[ \cD \ldots \right]$. Note that we define the energy in leads with respect to the
equilibrium dispersion, i.e., we do not include the external fields $U_{\alpha}$ and $\psi_{\alpha}$ in its definition, hence it represents the \emph{intrinsic} energy density.
Furthermore we define the heat current
\begin{align}
  Q_{\alpha}(t) & = (1 + \psi_{\alpha}) \left( J_{\alpha}(t) + U_{\alpha} I_{\alpha}(t) \right) ~, \label{Q}
\end{align}
which is the sum of the intrinsic energy current and the energy current due to the charge current in a potential multiplied by the TM field.

From the Heisenberg equation of motion, one obtains the well-known result that the currents are given in terms of the ``lesser'' Green's function connecting the leads and the junction,
\begin{subequations} \label{currents}
  \begin{align}
    I_{\alpha}(t) & = 2 \sum_{k, n} \RE\left[ V_{(\alpha, k), n} \mathcal{G}^{<}_{n, (\alpha, k)}(t, t) \right] ~, \label{I_G_lesser} \\
    J_{\alpha}(t) & = 2 \sum_{k, n} \bar\epsilon_{\alpha, k} \RE\left[ V_{(\alpha, k), n} \mathcal{G}^{<}_{n, (\alpha, k)}(t, t) \right] ~. \label{J_G_lesser}
  \end{align}
\end{subequations}

In the following, we restrict the discussion to a single impurity site labeled by ``${0}$'' and drop the summation over the energy levels of the junction
accordingly. Moreover, we are focusing on noninteracting electrons, which allows us to write
\begin{align}
  & \mathcal{G}^{<}_{0, (\alpha, k)}(t, t) \nn
  & = \hbar^2 \sum_{\lambda, \lambda'} \mathcal{G}^{\mathrm{R}}_{0, \lambda}(t, t_{0}) \mathcal{G}^{<}_{\lambda, \lambda'}(t_{0}, t_{0})
  \mathcal{G}^{\mathrm{A}}_{\lambda', (\alpha, k)}(t_{0}, t) ~, \label{noninteractingDecomposition}
\end{align}
with $\lambda$ and $\lambda'$ being composite indices that run over all leads \emph{and} the impurity site. All Green's functions appearing in Eq.\ \eqref{noninteractingDecomposition}
can be expressed in terms of a parent Green's function with a generic complex frequency argument ${z}$.\cite{DiVentra:08} There are three ``spatial'' types of Green's functions: Green's functions
that describe processes between leads and within a lead (diagonal part), $\mathcal{G}_{(\alpha, k), (\alpha' , k')}(z)$, Green's functions
describing processes between a lead and the impurity and vice versa,
${\mathcal{G}_{0, (\alpha, k)}(z)}$ and ${\mathcal{G}_{(\alpha, k), 0}(z)}$, and finally the Green's function describing processes within the impurity,
${\mathcal{G}_{0, 0}(z)}$. All of them would be trivial without the coupling between the leads and the impurity site.
It is natural to express these three types of Green's function defining an embedding self-energy for the impurity site, i.e.,
\begin{subequations} \label{embedding}
  \begin{align}
    g_{\alpha, k}(z) & = \frac{1}{z - \epsilon_{\alpha, k}} ~, \label{leadGFfree} \\
    \Sigma(z) & = \sum_{\alpha} \Sigma_{\alpha}(z) = \sum_{\alpha} \sum_{k} \left| V_{\alpha, k} \right|^2 g_{\alpha, k}(z) ~, \label{embeddingSE}
  \end{align}
\end{subequations}
where $g_{\alpha, k}(z)$ is the free propagator, or bare Green's function, of the leads and $\Sigma(z)$ is the embedding self-energy, given by the sum over the embedding self-energies
provided by each lead. ${\Sigma_{\alpha}(z)}$ encodes the decay from the impurity into lead $\alpha$. The three aforementioned Green's functions are given by
\begin{subequations} \label{GF_z}
  \begin{align}
    \mathcal{G}_{0,0}(z) & = \frac{1}{z - \left(\epsilon_{\mathrm{imp}} - \mu\right) - \Sigma(z)} ~, \label{impurityGF} \\
    \mathcal{G}_{(\alpha, k),0}(z) & = g_{\alpha, k}(z) V_{\alpha, k} \mathcal{G}_{0,0}(z) ~, \label{tunnelingGF_1} \\
    \mathcal{G}_{0,(\alpha, k)}(z) & = \mathcal{G}_{0,0}(z) V^\star_{\alpha, k} g_{\alpha, k}(z) ~, \label{tunnelingGF_2} \\
    \mathcal{G}_{(\alpha, k), (\alpha' , k')}(z) & = \delta_{\alpha \alpha'} \delta_{k k'} g_{\alpha, k}(z) \label{leadGF} \\
    & + g_{\alpha, k}(z) V_{\alpha, k} \mathcal{G}_{0,0}(z) V^\star_{\alpha', k'} g_{\alpha', k'}(z) ~. \nonumber
  \end{align}
\end{subequations}

Independent of their respective ``spatial'' type the advanced, retarded, and Matsubara Green's functions in the frequency domain
are given by ${\mathcal{G}^{\mathrm{A}}(\hbar \omega) = \mathcal{G}(\hbar \omega - i  \eta)}$,  ${\mathcal{G}^{\mathrm{R}}(\hbar \omega) = \mathcal{G}(\hbar \omega + i  \eta)}$
and  ${\mathcal{G}^{\mathrm{M}}(z_n) = \mathcal{G}(i  z_n)}$, respectively. $\eta$ is the usual positive infinitesimal enforcing the
advanced or retarded character of the Green's function and $z_n$ are the fermionic Matsubara frequencies given by ${z_n = \frac{\pi (2n + 1)}{\beta}}$ for integer $n$.
In the time domain, the Green's functions appearing in Eq.\ \eqref{noninteractingDecomposition} are
\begin{subequations} \label{GF_t}
  \begin{align}
    \mathcal{G}^{\mathrm{R}}(t, t_{0}) & = \frac{1}{2 \pi} \indll{\omega}{-\infty}{\infty} \mathcal{G}^{\mathrm{R}}(\hbar \omega) e^{- i  \omega(t - t_{0})} ~, \label{GFR_t_t0} \\
    \mathcal{G}^{\mathrm{A}}(t_{0}, t) & = \frac{1}{2 \pi} \indll{\omega}{-\infty}{\infty} \mathcal{G}^{\mathrm{A}}(\hbar \omega) e^{i  \omega(t - t_{0})} ~, \label{GFA_t0_t} \\
    \mathcal{G}^{<}(t_{0}, t_{0}) & = \frac{i}{\hbar \beta} \sum_{n = -\infty}^{\infty} \mathcal{G}^{\mathrm{M}}(z_n) e^{i  z_{n} \eta} ~, \label{GFM_t0}
  \end{align}
\end{subequations}
where the infinitesimal $\eta$ in Eq.\ \eqref{GFM_t0} ensures that we obtain the ``lesser'' Green's function. The summation over the Matsubara frequencies
is evaluated by the common contour integration technique using the Fermi-Dirac distribution $f(z) = \left(e^{\beta z} +1\right)^{-1}$, which has poles with residues $-\frac{1}{\beta}$
at the Matsubara frequencies. Deforming the contour encircling the Matsubara frequencies to run along the real frequency axis we get the well-known result
\begin{align}
  \mathcal{G}^{<}(t_{0}, t_{0}) & = \frac{1}{2 \pi \hbar} \indll{\epsilon}{-\infty}{\infty} f(\epsilon)
  \left( \mathcal{G}^{\mathrm{A}}(\epsilon) - \mathcal{G}^{\mathrm{R}}(\epsilon) \right) ~. \label{initial_occupations}
\end{align}
Although it appears that, by virtue of Eq.\ \eqref{initial_occupations}, we expressed Eq.\ \eqref{GFM_t0} in terms of the Green's functions given
in Eqs.\ \eqref{GFR_t_t0} and \eqref{GFA_t0_t} it is crucial to remember
that $\mathcal{G}^{<}(t_{0}, t_{0})$ represents the initial preparation of our system. This means that the dispersions entering in the definition of the bare Green's function and the
embedding self-energy determining $\mathcal{G}^{\mathrm{A/R}}(\epsilon)$ in Eq.\ \eqref{initial_occupations} are the \emph{unperturbed} dispersions $\bar\epsilon_{\alpha, k}$
defined in Eq.\ \eqref{equilibriumDispersion}, while the dispersions in Eqs.\ \eqref{GFR_t_t0} and \eqref{GFA_t0_t} are the \emph{perturbed} dispersions $\tilde\epsilon_{\alpha, k}$
of Eq.\ \eqref{propagationDispersion}. In order to keep track of this important difference, we rewrite Eq.\ \eqref{noninteractingDecomposition},
\begin{align}
  \mathcal{G}^{<}_{0, (\alpha, k)}(t, t) & = \hbar \sum_{\lambda, \lambda'}
  \frac{1}{2 \pi} \indll{\epsilon}{-\infty}{\infty} f(\epsilon) \left( \bar{\mathcal{G}}^{\mathrm{A}}_{\lambda, \lambda'}(\epsilon) - \bar{\mathcal{G}}^{\mathrm{R}}_{\lambda, \lambda'}(\epsilon) \right) \nn
  & \times \frac{1}{2 \pi} \indll{\omega}{-\infty}{\infty} e^{- i  \omega(t - t_{0})} \frac{1}{2 \pi} \indll{\omega'}{-\infty}{\infty} e^{i  \omega'(t - t_{0})} \nn
  & \times \tilde{\mathcal{G}}^{\mathrm{R}}_{0, \lambda}(\hbar \omega) \tilde{\mathcal{G}}^{\mathrm{A}}_{\lambda', (\alpha, k)}(\hbar \omega') ~, \label{G_lesser_0ak}
\end{align}
where we denote Green's functions that involve the equilibrium dispersions ${\bar{\epsilon}_{\alpha, k}}$ by $\bar{\mathcal{G}}$ and Green's function depending on ${\tilde{\epsilon}_{\alpha, k}}$
by ${\tilde{\mathcal{G}}}$. Equation \eqref{G_lesser_0ak} is the starting point for the calculation of the long-time limit.

\section{ Steady-state limit } \label{SEC:SteadyState}

In this section, we discuss the long-time limit of the particle and energy current given by Eqs.\ \eqref{currents}. The presented analysis
follows closely the derivation of Stefanucci and Almbladh.\cite{StefanucciAlmbladh:04}
The currents are given in terms of the ``lesser'' Green's function $\mathcal{G}^{<}_{0, (\alpha, k)}(t, t)$. From Eq.\ \eqref{G_lesser_0ak}, we can see that
for long times, $t \gg t_{0}$, the expression for $\mathcal{G}^{<}_{0, (\alpha, k)}(t, t)$ involves rapidly oscillating exponentials, which cancel any well-behaved
function (Riemann-Lebesgue theorem). Accordingly, in the long-time limit, the only non-vanishing terms arise from strongly peaked functions multiplying the
exponentials. It is instructive to consider the following simple example: Suppose we have a function ${\mathcal{F}(\omega)}$ with a simple pole in the
lower half of the complex frequency plane. We investigate the behavior of the integral
\begin{align}
  F(t-t_{0}) = \tfrac{1}{2 \pi} \indll{\omega}{-\infty}{\infty} e^{-i  \omega(t-t_{0})} \mathcal{F}(\omega) ~, \label{RiemannLebesgueExample_1}
\end{align}
for ${t \gg t_{0}}$. We can close the integration contour with a semi circle in the lower half of the complex frequency plane. Since the arc of the
semi circle does not contribute due to the exponential in Eq.\ \eqref{RiemannLebesgueExample_1}, we obtain simply
\begin{align}
  F(t-t_{0}) = - i e^{-i  \left(\omega_{0} - i  \frac{1}{\tau} \right) (t-t_{0})} \Res\left[\mathcal{F}(\omega_{0} - i  \tfrac{1}{\tau})\right] ~, \label{RiemannLebesgueExample_2}
\end{align}
where $\omega_{0} - i  \frac{1}{\tau}$ is the pole of ${\mathcal{F}(\omega)}$. Clearly, for ${t - t_{0} \gg \tau}$, the function $F(t-t_{0})$ vanishes exponentially.
However, for ${t - t_{0} \sim \tau}$, $F(t-t_{0})$ oscillates with frequency $\omega_{0}$.

The previous example helps to understand the regime of the long-time limit. We discard all poles of the Green's functions \emph{except}
for the poles due to bare Green's functions. The poles of the bare Green's functions are only infinitesimally, i.e., by $\pm i  \eta$, away from the real axis.
Since we only keep these poles, we are in the regime ${\tau \ll (t - t_{0}) \ll \frac{\hbar}{\eta}}$, where $\eta$ tends to zero.
The time scale $\tau$ is the time scale of the decay of electrons into the leads and therefore the long-time limit means that we are looking
at the system at a time much larger than typical relaxation time $\tau$. Note that we exclude the possibility of bound states
outside the continuum provided by the leads. If bound states would be present a steady state cannot be reached and the system would
oscillate with frequencies given by the energy differences associated with transitions between bound states and transitions between bound states and the occupations
edges of the continuum.\cite{KhosraviGross:08,KhosraviGross:09}

In Appendix \ref{APP:particleCurrent}, we derive the long-time limit of the particle current
\begin{align}
  I_{\alpha} & \equiv \lim_{t \to \infty} I_{\alpha}(t) = \frac{1}{\hbar} \sum_{\alpha'} \frac{1}{2 \pi} \indll{\epsilon}{-\infty}{\infty} f_{\alpha'} \nn
  & \times \frac{\Gamma_{\alpha'}(\epsilon)\Gamma(\epsilon) \delta_{\alpha \alpha'} - \Gamma_{\alpha'}(\epsilon)\Gamma_{\alpha}(\epsilon)}
  {\left(\epsilon - \left(\epsilon_{\mathrm{imp}} - \mu\right) - \frac{1}{2}\Lambda(\epsilon) \right)^2 + \left(\frac{1}{2}\Gamma(\epsilon)\right)^2} ~, \label{I_LTL_1}
\end{align}
where ${\Gamma_{\alpha}(\epsilon)}$ is twice the imaginary part and ${\Lambda_{\alpha}(\epsilon)}$ twice the real part of the (advanced) embedding self-energy ${\Sigma_{\alpha}(\epsilon)}$.
Since ${\Gamma(\epsilon) = \sum_{\alpha} \Gamma_{\alpha}(\epsilon)}$, it is straightforward to see from Eq.\ \eqref{I_LTL_1}
that ${\sum_{\alpha} I_{\alpha} = 0}$, expressing the fact that the particle current is conserved. An equivalent expression
for the steady-state current is given by
\begin{align}
  I_{\alpha} & = \frac{1}{\hbar} \sum_{\alpha'} \frac{1}{2 \pi} \indll{\epsilon}{-\infty}{\infty} \nn
  & \times \frac{\Gamma_{\alpha'}(\epsilon)\Gamma_{\alpha}(\epsilon) \left(f_{\alpha} - f_{\alpha'} \right)}
  {\left(\epsilon - \left(\epsilon_{\mathrm{imp}} - \mu\right) - \frac{1}{2}\Lambda(\epsilon) \right)^2 + \left(\frac{1}{2}\Gamma(\epsilon)\right)^2} ~. \label{I_LTL_2}
\end{align}
We stress that the embedding self-energy due to lead $\alpha$ (or equivalently ${\Gamma_{\alpha}(\epsilon)}$ and
${\Lambda_{\alpha}(\epsilon)}$) depends on the applied potentials $\psi_{\alpha}$ and $U_{\alpha}$ (cf.\ Eq.\ \eqref{Gamma_def} in Appendix \ref{APP:particleCurrent}).
Equation \eqref{I_LTL_2} highlights, however, that in the linear response regime, i.e., to first order in the biases, this dependence can be neglected,
because the difference in the occupation functions is already first order in $U_{\alpha}$ and $\psi_{\alpha}$.

The expression for the energy current is derived in Appendix \ref{APP:energyCurrent} and can be written in the following compelling form
\begin{align}
  J_{\alpha} & = \frac{1}{\hbar} \sum_{\alpha'} \frac{1}{2 \pi} \indll{\epsilon}{-\infty}{\infty} \left( \frac{\epsilon}{1+\psi_{\alpha}} - U_{\alpha} \right) f_{\alpha'} \nn
  & \times \frac{\Gamma_{\alpha'}(\epsilon)\Gamma(\epsilon) \delta_{\alpha \alpha'} - \Gamma_{\alpha'}(\epsilon)\Gamma_{\alpha}(\epsilon)}
  {\left(\epsilon - \left(\epsilon_{\mathrm{imp}} - \mu\right) - \frac{1}{2}\Lambda(\epsilon) \right)^2 + \left(\frac{1}{2}\Gamma(\epsilon)\right)^2} ~. \label{J_LTL}
\end{align}
Note that, in contrast to the particle current [cf.\ Eq.\ \eqref{I_LTL_1}], the energy current \eqref{J_LTL} is not conserved. 
However, the heat current, defined in Eq.\ \eqref{Q}, Sec.\ \ref{SEC:ThermoelectricTransport} is conserved. The energy current ${J_{\alpha}}$ is the \emph{intrinsic}
energy current while the conserved heat current $Q_{\alpha}$ also includes the applied potential and TM field.
The difference of $Q_{\alpha}$ and $J_{\alpha}$ is second order in the applied fields, i.e.,
\begin{align}
  Q_{\alpha} - J_{\alpha} = \psi_{\alpha} J_{\alpha} + (1 + \psi_{\alpha}) U_{\alpha} I_{\alpha} ~. \label{Q_J_diff}
\end{align}
We conclude by giving the heat current in the form similar to \ Eq.\ \eqref{I_LTL_2}, i.e.,
\begin{align}
  Q_{\alpha} & = \frac{1}{\hbar} \sum_{\alpha'} \frac{1}{2 \pi} \indll{\epsilon}{-\infty}{\infty} \epsilon \nn
  & \times \frac{\Gamma_{\alpha'}(\epsilon)\Gamma_{\alpha}(\epsilon) \left(f_{\alpha} - f_{\alpha'} \right)}
  {\left(\epsilon - \left(\epsilon_{\mathrm{imp}} - \mu\right) - \frac{1}{2}\Lambda(\epsilon) \right)^2 + \left(\frac{1}{2}\Gamma(\epsilon)\right)^2} ~. \label{Q_LTL}
\end{align}
Again, we see that, to first order in the biases, the dependence of the transmission function on the applied potentials $U_{\alpha}$ and $\psi_{\alpha}$
can be neglected. Furthermore, we note that in the linear regime the heat current $Q_{\alpha}$ and the energy current $J_{\alpha}$ are identical. This can
be seen from Eq.\ \eqref{Q_J_diff} since the currents themselves are already first order in the applied biases.

\section{ Landauer-B{\"u}ttiker versus Luttinger approach to thermal transport } \label{SEC:LBvsTM}

In the previous section, we have given the steady-state particle and energy/heat current employing Luttinger's idea of
the TM field ${\psi}$ as a proxy for temperature variations. We have found that, for noninteracting systems, the currents are
given by
\begin{subequations} \label{steadyStateCurrents}
  \begin{align}
    I_{\alpha} & = \frac{1}{\hbar} \sum_{\alpha'} \frac{1}{2 \pi} \indll{\epsilon}{-\infty}{\infty} T_{\alpha \alpha'}(\epsilon) \left(f_{\alpha} - f_{\alpha'}\right) ~, \label{ParticleCurrent} \\
    Q_{\alpha} & = \frac{1}{\hbar} \sum_{\alpha'} \frac{1}{2 \pi} \indll{\epsilon}{-\infty}{\infty} \epsilon T_{\alpha \alpha'}(\epsilon) \left(f_{\alpha} - f_{\alpha'}\right) ~, \label{HeatCurrent}
  \end{align}
\end{subequations}
in terms of the transmission function
\begin{align}
  T_{\alpha \alpha'}(\epsilon) & = \frac{\Gamma_{\alpha'}(\epsilon)\Gamma_{\alpha}(\epsilon)}
  {\left(\epsilon - \left(\epsilon_{\mathrm{imp}} - \mu\right) - \frac{1}{2}\Lambda(\epsilon) \right)^2 + \left(\frac{1}{2}\Gamma(\epsilon)\right)^2} ~. \label{transmissionFunction}
\end{align}
In the derivation (cf.\ Appendices \ref{APP:particleCurrent} and \ref{APP:energyCurrent}) we have seen that the transmission function depends
on the potentials ${U_{\alpha}}$ and on the TM fields $\psi_{\alpha}$ in the leads.
Now we compare the approach using Luttinger's TM field to the LB approach to thermal transport. The expressions for the 
particle and heat current in the LB approach are formally equivalent to Eq.\ \eqref{steadyStateCurrents}. However, in the LB approach, the only place
where the different temperatures of the leads enter is in the occupation factors:
\begin{subequations} \label{occupationFunctions}
  \begin{align}
    f_{\alpha}^{\mathrm{TM}} & = f_T\left(\frac{\epsilon}{1 + \psi_{\alpha}} - U_{\alpha}\right) \nn
    & = \left(\exp\left(\tfrac{1}{k_{\mathrm{B}} T} \left(\frac{\epsilon}{1 + \psi_{\alpha}} - U_{\alpha}\right) \right) + 1 \right)^{-1} ~, \label{occupationFunctionsTM} \\
    f_{\alpha}^{\mathrm{LB}} & = f_{T_{\alpha}}\left(\epsilon - U_{\alpha}\right) \nn
    & = \left(\exp\left(\tfrac{1}{k_{\mathrm{B}} T_{\alpha}} \left(\epsilon - U_{\alpha}\right) \right) + 1 \right)^{-1} ~. \label{occupationFunctionsLB}
  \end{align}
\end{subequations}
In Eq.\ \eqref{occupationFunctionsTM} the Fermi functions in the TM approach are shown, whereas in Eq.\ \eqref{occupationFunctionsLB} the Fermi functions in the LB approach are given.
First of all, we note that--to linear order in the biases--the difference of the occupation factors that enter in the expression for the currents [cf.\ Eqs.\ \eqref{steadyStateCurrents}]
are identical if we identify the variation $\delta \psi_{\alpha}$ in the TM approach with the relative temperature variation $\frac{T_{\alpha}- T}{T}$ in the LB approach ($T$ is
the reference temperature in both approaches), i.e.,
\begin{align}
  f_{\alpha} - f_{\alpha'} & \approx - f'(\epsilon) \big(\left(\delta U_{\alpha} - \delta U_{\alpha'} \right)
  + \epsilon \left(\delta \psi_{\alpha} - \delta \psi_{\alpha'} \right) \big) \nn
  & \approx - f'(\epsilon) \big( \delta U + \epsilon \delta \psi \big) ~, \label{occupationDifferenceLinearized}
\end{align}
where $\delta U$ is the potential difference between lead $\alpha$ and lead $\alpha'$ and $\delta \psi$ is the relative temperature difference between lead $\alpha$ and lead $\alpha'$.
This supports strongly the notion of the TM field $\psi$ as the mechanical ``proxy'' for relative temperature variations initiated by Luttinger.
Moreover, this implies that in the linear response regime the two approaches yield identical results, because the transmission functions are evaluated
at zero biases since the difference in the occupations is already first order in the applied fields.

The difference between the LB and the TM approach beyond the linear regime has two sources:
first of all, there is a difference in the occupation functions. We can rewrite the occupation function in the TM approach as
\begin{align}
  f_{\alpha}^{\mathrm{TM}} & = f_{T_{\alpha}}\left(\epsilon - (1 + \psi_{\alpha})U_{\alpha}\right) ~, \label{occupationFunctionTM_LB}
\end{align}
which differs from the occupation function in the LB approach, $f_{T_{\alpha}}\left(\epsilon - U_{\alpha}\right)$, by a simple rescaling of the
potential bias by $(1 + \psi_\alpha)$. The effective bias in the TM approach is increased for higher temperatures ($\psi >0$) and decreased for lower
temperatures ($\psi < 0$).
Secondly, there is a difference in the transmission function. In order to analyze this difference we focus on the numerator of Eq.\ \eqref{transmissionFunction}, which is given by
the product of the decay rates.\footnote{Strictly speaking $\Gamma_{\alpha}(\epsilon)/(2 \hbar)$ is the decay rate into lead $\alpha$ or the
  inverse lifetime of the impurity state due to the presence of lead $\alpha$.} They are given by
\begin{subequations} \label{broadenings}
  \begin{align}
    \frac{\Gamma^{\mathrm{TM}}_{\alpha}}{2 \pi}
    & = \sum_k \left| V_{\alpha, k} \right|^2\delta\big(\epsilon - (1 + \psi_{\alpha}) \left(\bar{\epsilon}_{\alpha, k} + U_{\alpha}\right) \big) ~, \label{broadeningTM} \\
    \frac{\Gamma^{\mathrm{LB}}_{\alpha}}{2 \pi}
    & = \sum_k \left| V_{\alpha, k} \right|^2\delta\big(\epsilon - \left(\bar{\epsilon}_{\alpha, k} + U_{\alpha}\right) \big) ~, \label{broadeningLB}
  \end{align}
\end{subequations}
which are the density of states weighted by the hopping probability $\left| V_{\alpha, k} \right|^2$. Comparing Eqs.\ \eqref{broadeningTM}
and \eqref{broadeningLB}, we see that the density of states in the TM approach is effectively stretched for elevated temperatures (${\psi>0}$) and squeezed for
lowered temperatures (${\psi<0}$). Ignoring the effect of the denominator in the transmission function, we can qualitatively discuss the
differences in particle and heat transport in the TM approach compared to the LB approach. This is justified because the denominator
is a strictly positive function and hence it cannot change the behavior of the integrand qualitatively.

\begin{figure}
  \includegraphics[width=.45\textwidth]{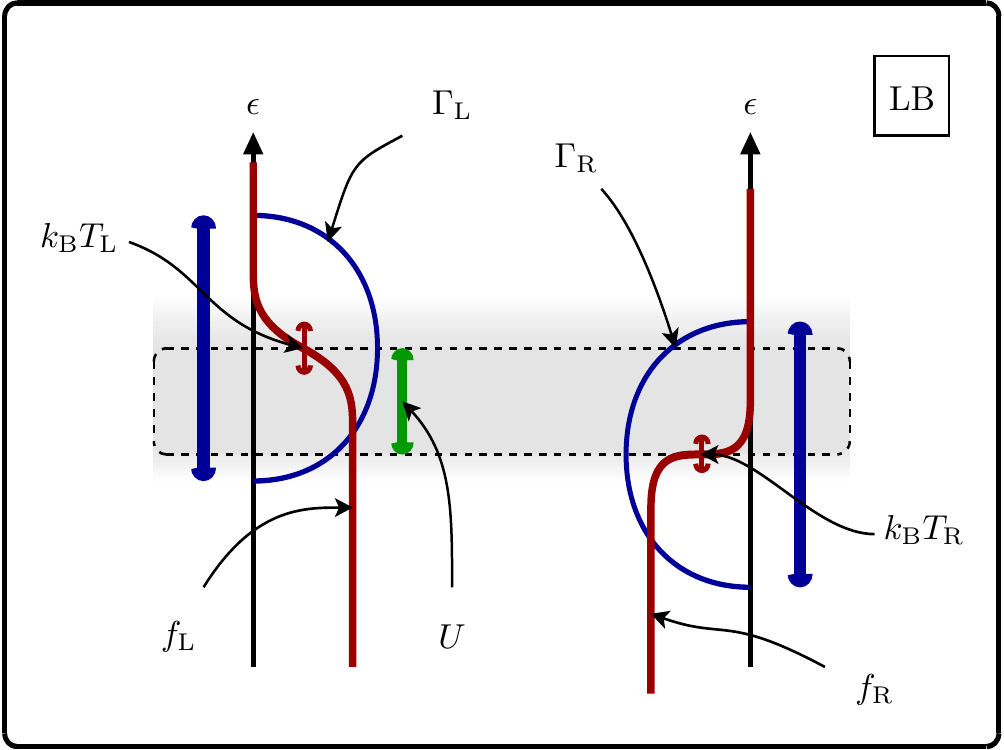}
  \includegraphics[width=.45\textwidth]{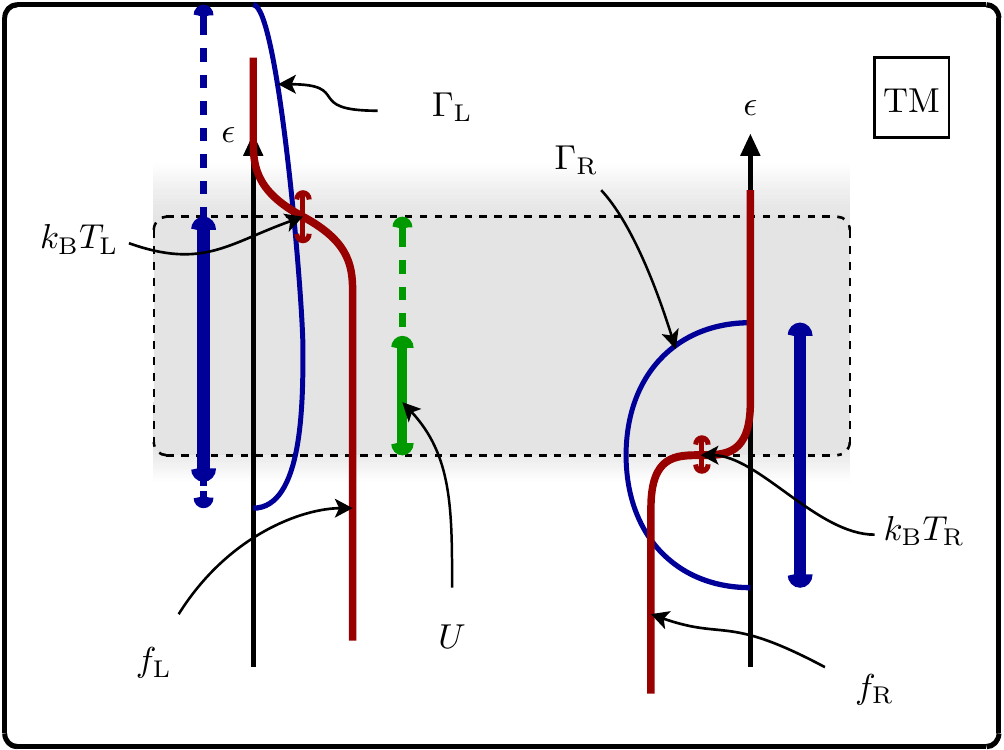}
  \caption{(Color online) Sketch of a typical transport scenario between two leads. The upper panel depicts the
    situation in the LB approach and the lower panel in the TM approach. A potential bias $U$ is applied to the left lead
    and the temperature in the left lead is raised. Note that in the TM approach the effective potential, determining
    the position of the occupation function $f_{\mathrm{L}}$, is rescaled by $(1 + \psi) = \tfrac{T_{\mathrm{L}}}{T}$. This is indicated by
    the dashed extension of the range (green) representing the applied bias $U$. Furthermore, the band of the left lead is rescaled by $(1 + \psi)$
    as shown by the dashed extension of the original bandwidth (blue). \label{FIG:transportSketch}}
\end{figure}

Figure \ref{FIG:transportSketch} compares the expression for the currents in the LB and the TM approach schematically. The upper panel shows the LB approach
where the leads are held initially at different temperatures and at $t=0$ a potential bias $U$ is applied to the left lead. The application of the potential bias results in a simultaneous
shift of the band and the occupation function. This opens a transport window (gray-shaded region) roughly the size of the applied bias. Within the transport window,
we have an excess of electrons on the left side and hence a particle current flows to the right. The effect of having finite temperatures in the leads is to soften
the transport window by $k_{\mathrm{B}} T_{\alpha}$. In the presented case, we have an elevated temperature in the left lead implying an enhanced softening on the upper
edge of the transport window which affects the currents.
The lower panel shows the situation in the TM approach where the leads are initially held at the same temperature and at $t=0$ a potential bias $U$ and a TM field $\psi$
are switched in the left lead. Again, the band is shifted upwards by $U$, but in contrast to the LB approach the band is stretched due to the positive TM field.
Furthermore, the occupation function is shifted by $(1+\psi)U$, i.e., by a rescaled potential. In the case of a positive TM field, this opens a transport window which is bigger than the
transport window in the LB approach. By itself the increased transport window should increase the currents. The stretching of the band, however, reduces the
number of available states in the transport window, which opposes the increase of the currents. In general, one cannot determine whether the effect of an increased transport
window dominates the decrease in available electrons in this transport window or vice versa.

Before we discuss explicit examples comparing the two approaches, we recall that the LB approach is based, by its very construction, on a partitioning of the system.\cite{DiVentra:08}
The transport setup is initially uncontacted, i.e., the initial state state is described by a density matrix
\begin{align}
  \cD^{\mathrm{LB}} & = \frac{\prod_{\alpha}\exp^{ - \beta_{\alpha} \left(\cH_{\alpha} - \mu_{\alpha} \cN_{\alpha} \right) }}{\mathcal{Z}^{\mathrm{LB}}} ~, \label{equilibriumDM_LB}
\end{align}
which, in contrast to the initial density matrix in the TM approach [cf.\ Eq.\ \eqref{equilibriumDM}], is defined with respect to the individual
Hamiltonians $\cH_{\alpha}$ of the leads. 
The LB approach requires that the coupling Hamiltonian $\sum_{\alpha}\cH_{\alpha, \mathrm{imp}}$ is ignored in the initial preparation 
in order to assign a specific temperature to each lead.
Each lead subsystem is initially coupled to its own reservoir. The initial density matrix in the TM approach, however, is determined by coupling the
entire system of leads and impurity to a single reservoir. As it turns out, it is possible to mimic the LB approach by applying the TM field
in the preparation of the initial state. This means that instead of switching the TM field ``on'' at the beginning of the propagation we prepare
the system in the presence of the TM field, and turn it ``off'' at $t=0$. The analysis for the steady-state currents, presented in Sec.\ \ref{SEC:SteadyState},
remains valid. The only difference is that we now have an equilibrium lead dispersion
\begin{align}
  \bar\epsilon_{\alpha, k} = \left(1 + \psi_{\alpha}\right) \left( \epsilon_{\alpha, k} - \mu \right) ~, \label{equilibriumDispersionLB}
\end{align}
which is modified by the TM field while the dispersion during the time-propagation is given by
\begin{align}
  \tilde\epsilon_{\alpha, k} = \epsilon_{\alpha, k} - \mu + U_{\alpha} ~. \label{propagationDispersionLB}
\end{align}
In this way, we reproduce exactly the expression of the LB approach via the TM field provided we relate $T_{\alpha} = T / (1 + \psi_{\alpha})$.
Accordingly, we have to identify $\delta T_{\alpha} / T = - \delta \psi$ in the linear regime, whereas before we identified $\delta T_{\alpha} / T = \delta \psi$ (cf.\
Eq.\ \eqref{occupationFunctions} and subsequent discussion). In Ref.\ \onlinecite{Shastry:09} (cf.\ footnote p.\ 9), Shastry humorously referred to this sign reversal
as ``booby trap.'' Here, we point out that this is simply due to the fact that switching ``off'' a mechanical field is identical to switching ``on'' a mechanical field
in the opposite direction in the linear regime. Finally, we point out a caveat concerning gauge invariance. Usually, gauge invariance
implies that the currents do not change under a constant shift of all bias potentials, $U_\alpha \to U_\alpha + \Phi$. \footnote{Note that the
potential shift needs to be applied also to the junction.} Since the potentials are effectively rescaled by $1 + \psi_\alpha$ when the TM field is switched ``on,''
the corresponding gauge transformation is $U_\alpha \to U_\alpha + \Phi / (1 + \psi_\alpha)$.

In conclusion, we see that the TM field allows us to exactly reproduce the LB approach commonly employed to study thermal transport. Furthermore,
since the TM field enters as a mechanical field in the Hamiltonian, we can study the time-dependent situation in which the TM field is switched on,
which is outside the realm of the traditional LB approach. Hence the TM field is--in the presented sense--an extension of the statistical mechanical
temperature to a spatially and temporally varying driving field.

\section{ Symmetric leads weakly coupled to an impurity } \label{SEC:Impurity}

\begin{figure}
  \includegraphics[width=.48\textwidth,clip]{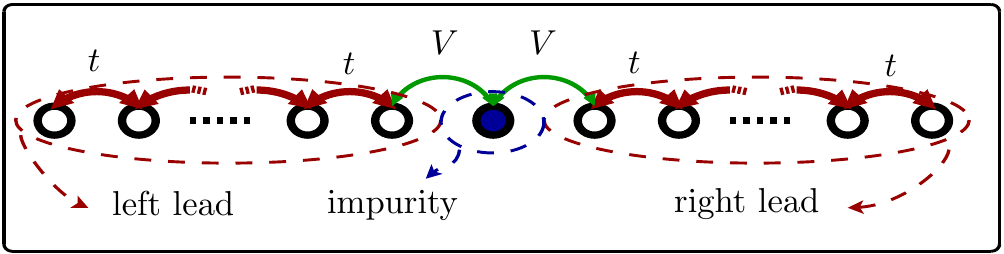}
  \caption{(Color online) Sketch of the Hamiltonian employed in the numerical examples. The left and right leads are
    modeled by infinite tight-binding chains which are characterized by the nearest neighbor hopping amplitude $t$.
    The hopping to the impurity site is described by the amplitude $V=0.1t$. The energy of the impurity site is at $\epsilon_{\mathrm{imp}} = \mu = 0$. \label{FIG:TBModel}}
\end{figure}
As an explicit example, we are considering two symmetric metallic leads (${\alpha = \mathrm{L}, \mathrm{R}}$)
weakly coupled to an impurity. Employing the model Hamiltonian, we have introduced in Sec.\ \ref{SEC:ThermoelectricTransport},
we set $t_{\alpha} / V_{\alpha} = 10$, for both the left ($\alpha = \mathrm{L}$) and the right (${\alpha = \mathrm{R}}$) lead.
The chemical potential is chosen to be at the center of the bands (${ \mu = 0}$), the impurity level is in resonance
(${\epsilon_{\mathrm{imp}} = 0}$) and the initial temperature is $\beta^{-1} = k_{\mathrm{B}} T = 0.1 t = V$. Biases are only applied to the left lead
unless specified otherwise. A schematic sketch of the employed model Hamiltonian is shown in Fig.\ \ref{FIG:TBModel}.

\subsection{ Particle current } \label{SEC:particleCurrent}
\begin{figure}
  \includegraphics[width=.48\textwidth,clip]{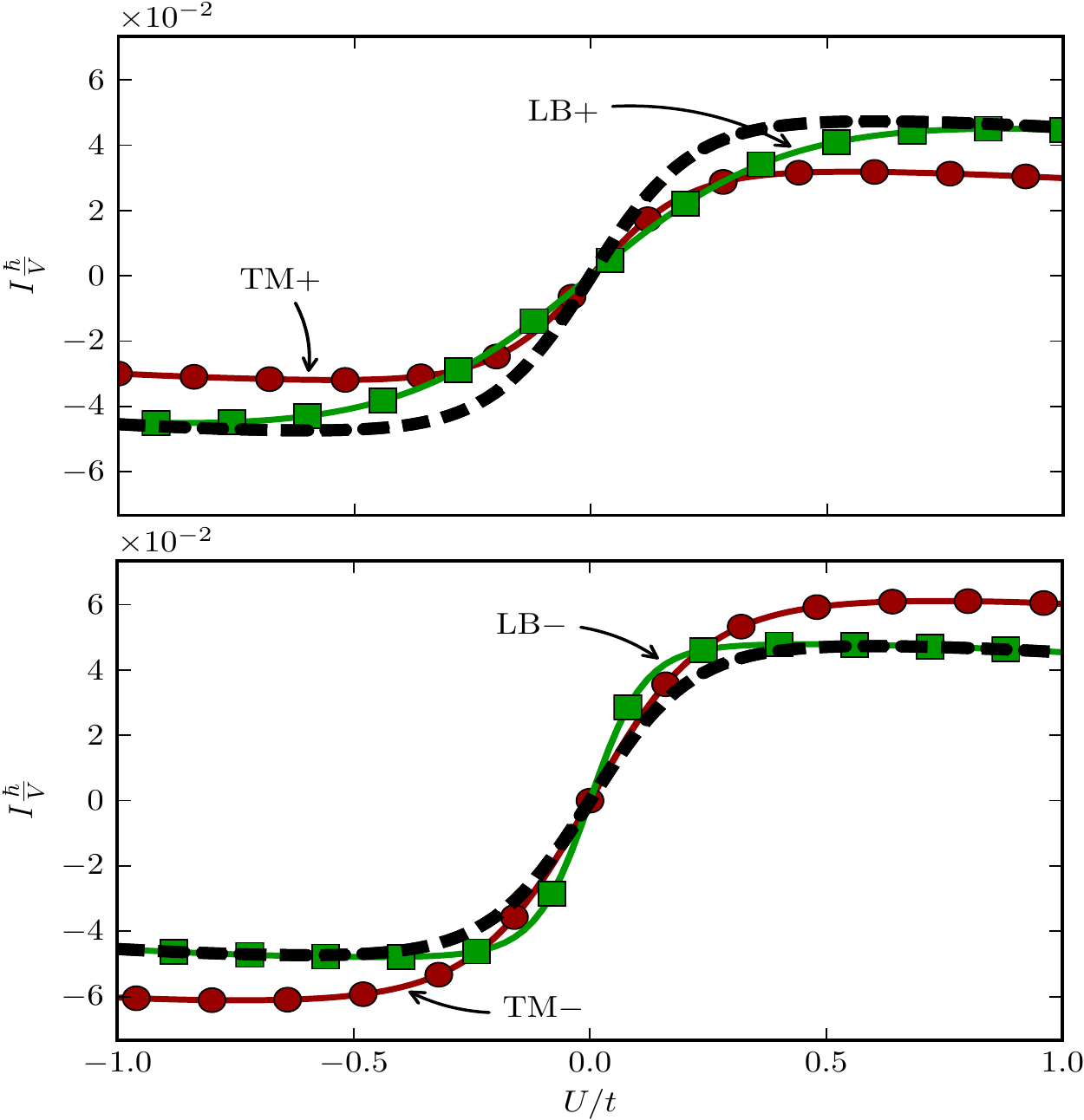}
  \caption{(Color online) Comparison of the steady-state particle currents ${I}$ in the TM and the LB approach. The currents are plotted against the potential bias
    $U$. The upper panel depicts the currents when the temperature in the left lead is raised to twice its initial value, i.e., $\psi = \delta T / T = 1$. The lower panels shows the currents
    when the temperature in the left lead is reduced to half its initial value ($\psi = \delta T / T = - 0.5$). The dashed, black curve shows the particle current at zero temperature
    difference for comparison. It is identical in the LB and the TM approach. The circles (red curve, labeled ``TM$\pm$'') correspond to the current in the TM approach and the squares
    (green curve, labeled ``LB$\pm$'') to the LB approach. \label{FIG:impurityParticleCurrentU}}
\end{figure}
Figure \ref{FIG:impurityParticleCurrentU} shows the comparison of the particle current
in the LB and the TM approach. The particle current is shown as function of the potential bias.
The upper panel depicts the currents for $\psi = \delta T / T = 1$, which corresponds to a temperature in the left lead that is
elevated to twice the temperature in the right lead. The lower panel shows the currents for $\psi = \delta T / T = -0.5$, which means
that the temperature in the left lead is lowered to half the temperature in the right lead.
As a reference we also show the current at zero TM field (dashed, black curve), i.e., at constant temperature throughout the device.
The sign of the particle current $I$ follows the sign of the applied bias voltage $U$. We can see that
raising the temperature leads to a reduction of the particle current in both approaches. Conversely lowering
the temperature has the opposite effect, i.e., the current increases. As mentioned earlier, the temperature of the
leads softens the transport window (cf.\ Fig.\ \ref{FIG:transportSketch}), which
means that in a region $k_{\mathrm{B}} T_{\alpha}$ we have partially occupied states. In combination with the specific shape
of the decay rate $\Gamma_{\alpha}(\epsilon)$ [cf.\ Eqs.\ \eqref{broadenings}], this leads to a reduced particle current 
compared to \emph{zero} temperature. Note, however, that this effect vanishes for larger potential biases in the LB approach,
indicating that the softening of the transport window can be neglected in this case. In view of the general discussion, presented in Sec.\ \ref{SEC:LBvsTM},
we conclude that the rescaling of the density of states dominates over the rescaling of the potential for large potential biases in the TM approach.
Furthermore, this effect does not disappear, i.e., we have a reduction of the current for elevated temperatures and an increase in current for lowered
temperatures even for large $U$. Conversely, for $|U| \lesssim 0.2$, there is less reduction of the current for elevated temperatures and less increase of the
current for lowered temperatures in the TM approach compared to the LB approach. This indicates that for small biases the effect of rescaling the transport window
dominates over the rescaling of the density of states.

\subsection{ Heat current } \label{SEC:heatCurrent}
\begin{figure}
  \includegraphics[width=.48\textwidth]{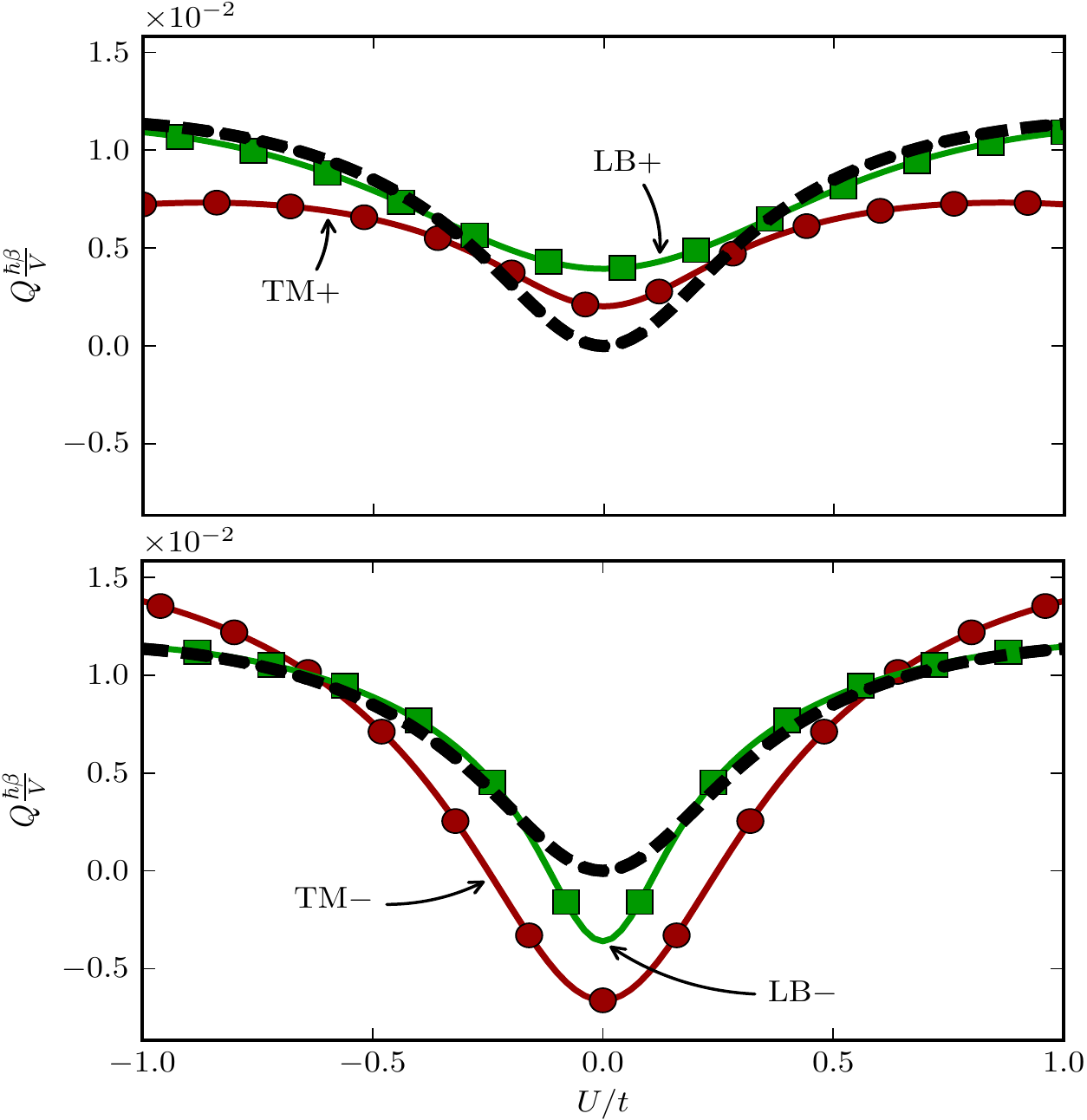}
  \caption{(Color online) Comparison of the steady-state heat currents ${Q}$ in the TM and the LB approach. The heat currents are plotted against the potential bias
    $U$. The upper panel depicts the heat currents when the temperature in the left lead is raised to twice its initial value ($\psi = 1$) and the lower panel shows the heat currents
    when the temperature in the left lead is reduced to half its initial value ($\psi = -0.5$). The dashed, black curve is the heat current at a uniform temperature throughout the system.
    The circles (red curve, labeled ``TM$\pm$'') correspond to the heat current in the TM approach and the squares (green curve, labeled ``LB$\pm$'')
    to the LB approach. \label{FIG:impurityHeatCurrentU}}
\end{figure}
In Fig.\ \ref{FIG:impurityHeatCurrentU}, we compare the heat current as a function of the applied potential bias for the LB and TM approaches.
In the upper panel of this figure the heat current is plotted against $U / t$ for $\psi = \delta T / T = 1$, whereas in the lower panel it is plotted
for $\psi = \delta T / T = -0.5$. The dashed, black line represents the heat current for equal temperatures in the leads: by construction, this is
the same regardless of whether we use the LB or the TM approach. We note that in the case of zero temperature difference the heat current
$Q$ has a unique direction independent on the applied potential bias $U$. This is due to the fact that the energy is measured with respect to the chemical potential: while for positive
$U$ electrons \emph{above} the chemical potential move from left to right, for negative $U$ electrons \emph{below} the chemical potential move from right to left. Accordingly,
the heat current does not change sign while the particle current does. Note that this implies a vanishing Peltier coefficient in the linear regime.
We return to this point when we are discussing the dependency of the heat and particle currents on a temperature bias.   

Now we consider a difference in temperature between the leads. For small biases, i.e., $|U| \lesssim 0.2 t$, we get a heat current that flows from the
``hotter'' to the ``colder'' lead. Hence we see that around $U\sim 0$ the heat current for a lowered temperature in the left lead (lower panel, Fig.\ \ref{FIG:impurityHeatCurrentU})
is negative in the TM and LB approach indicating that heat flows from the ``hotter'' right lead to the ``colder'' left lead. Similarly, for an elevated temperature
(upper panel, Fig.\ \ref{FIG:impurityHeatCurrentU}) the heat flows from the ``hotter'' left lead to the ``colder'' right lead. For larger biases, however, the heat current
is dominated by the applied potential bias and heat flows from left to right as in the situation where no temperature difference between the leads is present.
Note that this is not in contradiction to thermodynamic principles since applying a potential bias means that we perform work on the system and, hence, heat may flow from
the ``colder'' to the ``hotter'' lead.
Similar to the particle current, the heat current is reduced, relative to the equal temperatures case, when the temperature in the left lead is higher than the temperature
in the right lead in the region where the heat current is dominated by the contribution due to the potential bias.
Comparing the TM to the LB approach we find that, in the case of an elevated temperature in the left lead, the heat current in the TM approach is smaller than
the heat current in the LB approach in the whole range of $U$.

When the temperature in the left lead is lower than in the right lead, we also see qualitatively the same behavior in heat and particle currents, i.e., both currents
are increased compared to the currents at zero temperature difference for small ($|U| \lesssim 0.1 t $) and large ($|U| \gtrsim 0.5 t$) potential bias.
In the intermediate region, the heat current changes its direction. The specific potential bias range for which the heat current flows in its ``natural'' direction,
i.e., from ``hot'' to ``cold,'' differs in the two approaches. This can be understood from the fact that we have an effectively rescaled potential bias
in the TM approach. The change of direction in the heat current due to the applied potential happens at higher biases since $U$ is reduced by a factor $\sim 2$ for $\psi=-0.5$.

\begin{figure}
  \includegraphics[width=.48\textwidth]{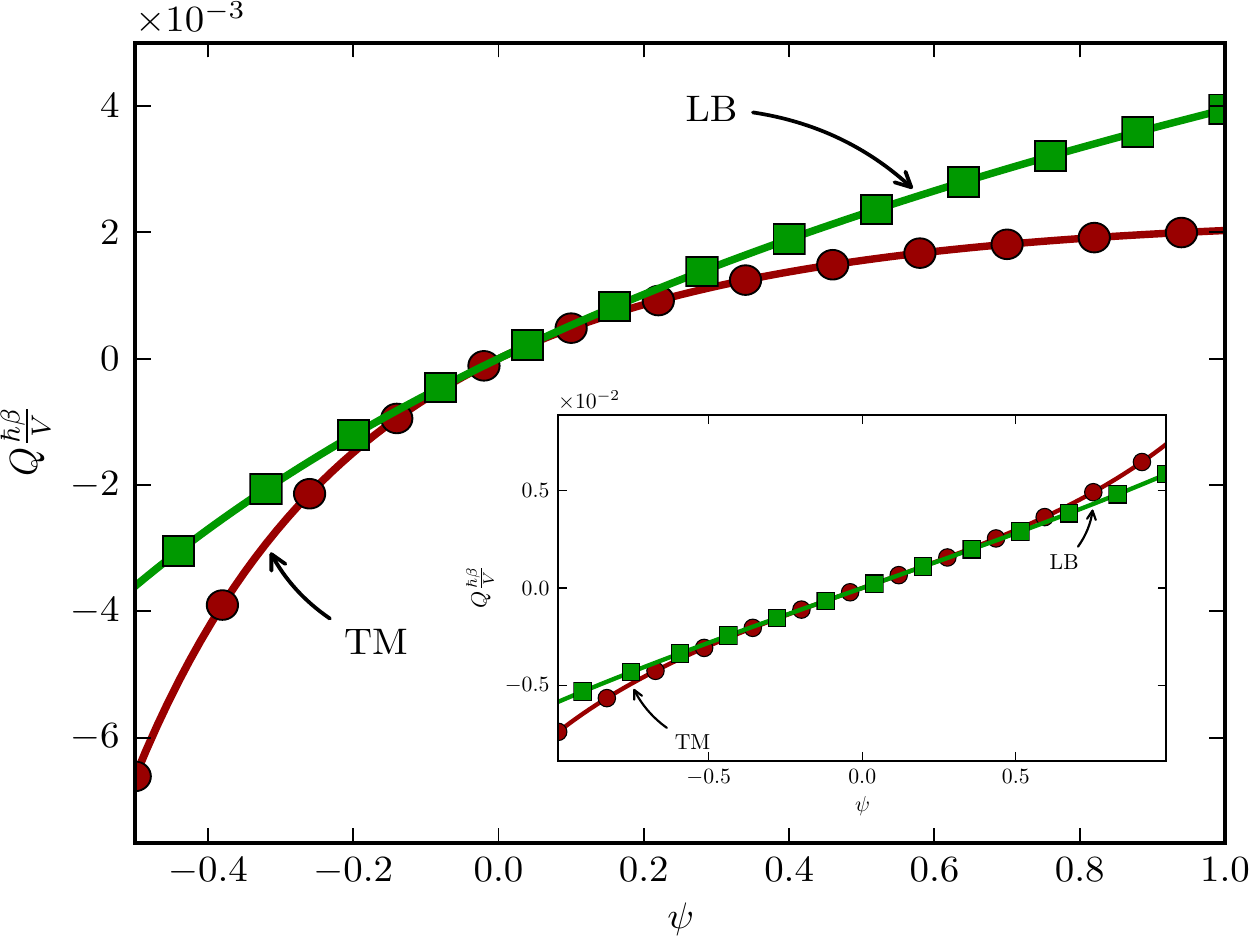}
  \caption{(Color online) Comparison of the steady-state heat current ${Q}$ in the TM and the LB approach at zero potential bias as a function
    of the relative temperature difference. The circles (red curve, labeled ``TM'') and squares (green curve, labeled ``LB'') show the heat current in the TM and LB approaches, respectively.
    While in the main plot the temperature is only changed in the left lead, the inset shows the heat current when the TM field (or relative temperature difference)
    is applied antisymmetrically in the left and right leads, i.e., $\psi_{\mathrm{L}} = \delta T_{\mathrm{L}} / T = - \psi_{\mathrm{R}} = - \delta T_{\mathrm{R}} / T$.
    In the inset we have defined $\psi = \psi_{\mathrm{L}} - \psi_{\mathrm{R}}$. $Q$ at the left boundary ($\psi = - 0.5$) of the main plot corresponds to $Q$
    at $U=0$ in the lower panel of Fig.\ \ref{FIG:impurityHeatCurrentU} and $Q$ at the right boundary ($\psi = 1$) of the main plot to $Q$ at $U=0$ in the upper panel.
    \label{FIG:impurityHeatCurrentPsi}}
\end{figure}
Next, in order to focus on the effect of a temperature or TM field difference between the left lead and the right lead, we investigate the currents at zero potential bias.
Since the chemical potential is at the center of the band for both leads, the situation is particle-hole symmetric. Together with the fact that the impurity site
is in perfect resonance, i.e., $\epsilon_{0} = \mu$, this implies that there is no particle current induced if we only apply a temperature gradient. Nevertheless, there
is heat transfer between the two leads as shown in Fig.\ \ref{FIG:impurityHeatCurrentPsi}. The asymmetry around $\psi = 0$ is due to the fact that we are only applying the temperature
difference in the left lead. The asymmetry disappears if TM fields of equal magnitude and opposite signs are applied to the two leads (cf.\ inset of Fig.\  \ref{FIG:impurityHeatCurrentPsi}).
As already seen in Fig.\ \ref{FIG:impurityHeatCurrentU}, the heat current in the TM approach is reduced compared to the LB approach
for elevated temperatures ($\psi > 0$) and increased for lowered temperatures ($\psi < 0$). In order to further investigate this we analyze the expression for the
heat current, Eq.\ \eqref{HeatCurrent}, for the case of a vanishing potential bias. First of all, we note that the occupation functions are identical in this case. Secondly, from
the definition of the decay rates [cf.\ Eqs.\ \eqref{broadenings}], we can see that
\begin{align}
  \Gamma^{\mathrm{TM}}_{\alpha}(\epsilon) = \frac{1}{1+\psi_\alpha}\Gamma^{\mathrm{LB}}_{\alpha}\left(\tfrac{\epsilon}{1+\psi_\alpha}\right) ~. \label{gammaConnection}
\end{align}
Ignoring the denominator of the transmission function for a moment--as we have done for the general discussion in Sec.\ \ref{SEC:LBvsTM}--we obtain for the heat current
in the TM approach,
\begin{align}
  Q^{\mathrm{TM}} & = \frac{1}{\hbar} \frac{1}{2 \pi} \indll{\epsilon}{-\infty}{\infty} \epsilon \Gamma^{\mathrm{TM}}_{\mathrm{L}}(\epsilon) \Gamma^{\mathrm{TM}}_{\mathrm{R}}(\epsilon)
  \left(f_{\mathrm{L}} - f_{\mathrm{R}}\right) \nn
  & = \frac{1}{\hbar} \frac{1}{2 \pi} \indll{\epsilon}{-\infty}{\infty} \frac{\epsilon}{1+\psi} \Gamma^{\mathrm{LB}}\left(\tfrac{\epsilon}{1+\psi}\right) \Gamma^{\mathrm{LB}}(\epsilon) \nn
  & {} \times \left( f_T\left(\tfrac{\epsilon}{1+\psi}\right) - f_T(\epsilon) \right) \nn
  & = \frac{1}{\hbar} \frac{1}{2 \pi} \indll{\epsilon}{-\infty}{\infty} \epsilon \Big( (1+\psi) \Gamma^{\mathrm{LB}}(\epsilon) \Gamma^{\mathrm{LB}}(\epsilon(1+\psi)) \nn
  & {} - \frac{1}{1+\psi} \Gamma^{\mathrm{LB}}\left(\tfrac{\epsilon}{1+\psi}\right) \Gamma^{\mathrm{LB}}(\epsilon) \Big) f_{T}(\epsilon) ~, \label{HeatCurrentLRTM}
\end{align}
where we first use relation \eqref{gammaConnection} and the explicit form of the occupation function [cf.\ Eqs.\ \eqref{occupationFunctions}] for a TM field $\psi$
applied to the left lead and subsequently shift the integration variable $\epsilon \to (1+\psi) \epsilon$ in the first term. Similarly we can write the heat current in the LB approach as
\begin{align}
  Q^{\mathrm{LB}} & = \frac{1}{\hbar} \frac{1}{2 \pi} \indll{\epsilon}{-\infty}{\infty} \epsilon \Gamma^{\mathrm{LB}}_{\mathrm{L}}(\epsilon) \Gamma^{\mathrm{LB}}_{\mathrm{R}}(\epsilon)
  \left(f_{\mathrm{L}} - f_{\mathrm{R}}\right) \nn
  & = \frac{1}{\hbar} \frac{1}{2 \pi} \indll{\epsilon}{-\infty}{\infty} \epsilon \Gamma^{\mathrm{LB}}(\epsilon) \Gamma^{\mathrm{LB}}(\epsilon)
  \left(f_T\left(\tfrac{\epsilon}{1+\psi}\right) - f_T(\epsilon)\right) \nn
  & = - \frac{1}{\hbar} \frac{1}{2 \pi} \indll{\epsilon}{-\infty}{\infty} \epsilon f_{T}(\epsilon) \Big( \Gamma^{\mathrm{LB}}(\epsilon) \Gamma^{\mathrm{LB}}(\epsilon) \nn
  & {} - (1+\psi)^2 \Gamma^{\mathrm{LB}}(\epsilon(1+\psi)) \Gamma^{\mathrm{LB}}(\epsilon(1+\psi)) \Big) ~. \label{HeatCurrentLRLB}
\end{align}
From the explicit form of the decay rate for our model, i.e.,
\begin{align}
  \Gamma_\alpha^{\mathrm{LB}}(\epsilon) = 2 \theta\left(1-\left(\tfrac{\epsilon}{2 t_\alpha}\right)^2\right) \frac{|V_\alpha|}{t_\alpha}
  \sqrt{1 - \left(\tfrac{\epsilon}{2 t_\alpha}\right)^2} ~, \label{decayRate}
\end{align}
where $\theta(x)$ is the Heaviside step function, one can show that the difference between the heat current in the TM and the LB approach goes as
\begin{align}
  Q^{\mathrm{TM}} - Q^{\mathrm{LB}} \approx A \psi^2 + \mathcal{O}(\psi^3) ~. \label{heatCurrentDifference1}
\end{align}
From Fig.\ \ref{FIG:impurityHeatCurrentPsi} we see that $A<0$.

The inset of Fig.\ \ref{FIG:impurityHeatCurrentPsi} shows the heat current when the relative temperature or TM field bias is applied antisymmetrically
in the left and right lead. By its very construction this situation is completely left-right antisymmetric. It is straight forward to perform an analysis
analogous to Eqs.\ \eqref{HeatCurrentLRTM} and \eqref{HeatCurrentLRLB} for the situation of an antisymmetrically applied TM field. We find that the difference
of the heat currents for small TM fields behaves as
\begin{align}
  Q^{\mathrm{TM}} - Q^{\mathrm{LB}} \approx B \psi^3 + \mathcal{O}(\psi^5) ~, \label{heatCurrentDifference2}
\end{align}
consistent with the fact that the setup is completely antisymmetric.
Equations \eqref{heatCurrentDifference1} and \eqref{heatCurrentDifference2} demonstrate that in both scenarios shown in Fig.\ \ref{FIG:impurityHeatCurrentPsi}
the heat currents from the TM and the LB approach are identical in the linear regime. Furthermore Eq.\ \eqref{heatCurrentDifference2} explains why the region
of validity for the linear approximation appears to be much larger in the scenario of an antisymmetrically applied temperature bias.

\section{ Local temperature } \label{SEC:localTemperature}

\begin{figure}
  \includegraphics[width=.48\textwidth,clip]{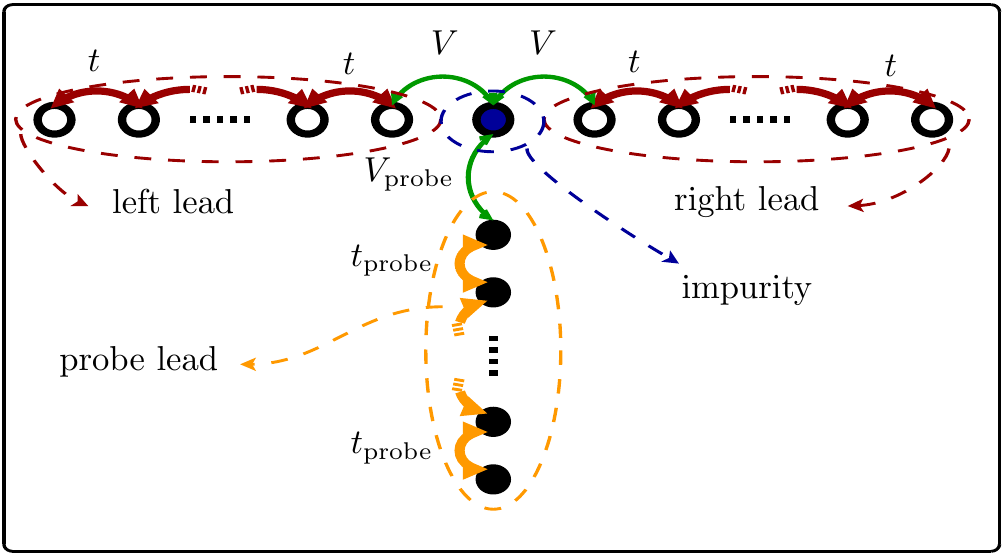}
  \caption{(Color online) Sketch of the Hamiltonian employed in the numerical computation of an effective local temperature.
    In addition to the Hamiltonian sketched in Fig.\ \ref{FIG:TBModel}, a probe lead is connected to the impurity.
    The probe lead is modeled by a half-filled tight-binding chain characterized by a hopping amplitude $t_\probe \gg t$
    and the coupling to the impurity site is described by the amplitude $V_{\probe} \ll V$. This implies that the probe lead can be treated in the
    wide-band limit. \label{FIG:TBModelProbe}}
\end{figure}
Finally, we address the definition of a local temperature.\cite{EngquistAnderson:81,CasoLozano:10,DubiDiVentra:11,BergfieldStafford:13} The concept of
a local effective temperature has come into focus due to the fact that
nowadays scanning thermal microscopy experiments achieve spatial resolution in the
nanometer range.\cite{Majumdar:99,YuKim:11,KimLee:11,KimReddy:12,MengesGotsmann:12}
A common procedure to address the idea of a local temperature from the theoretical
side is to mimic the experimental setup by introducing a (metallic) tip that is
weakly coupled to the nanoscale device under investigation. The potential and temperature
bias in the tip are then chosen to yield zero particle and heat current.\cite{DiVentra:08}

\begin{figure}
  \includegraphics[width=.48\textwidth]{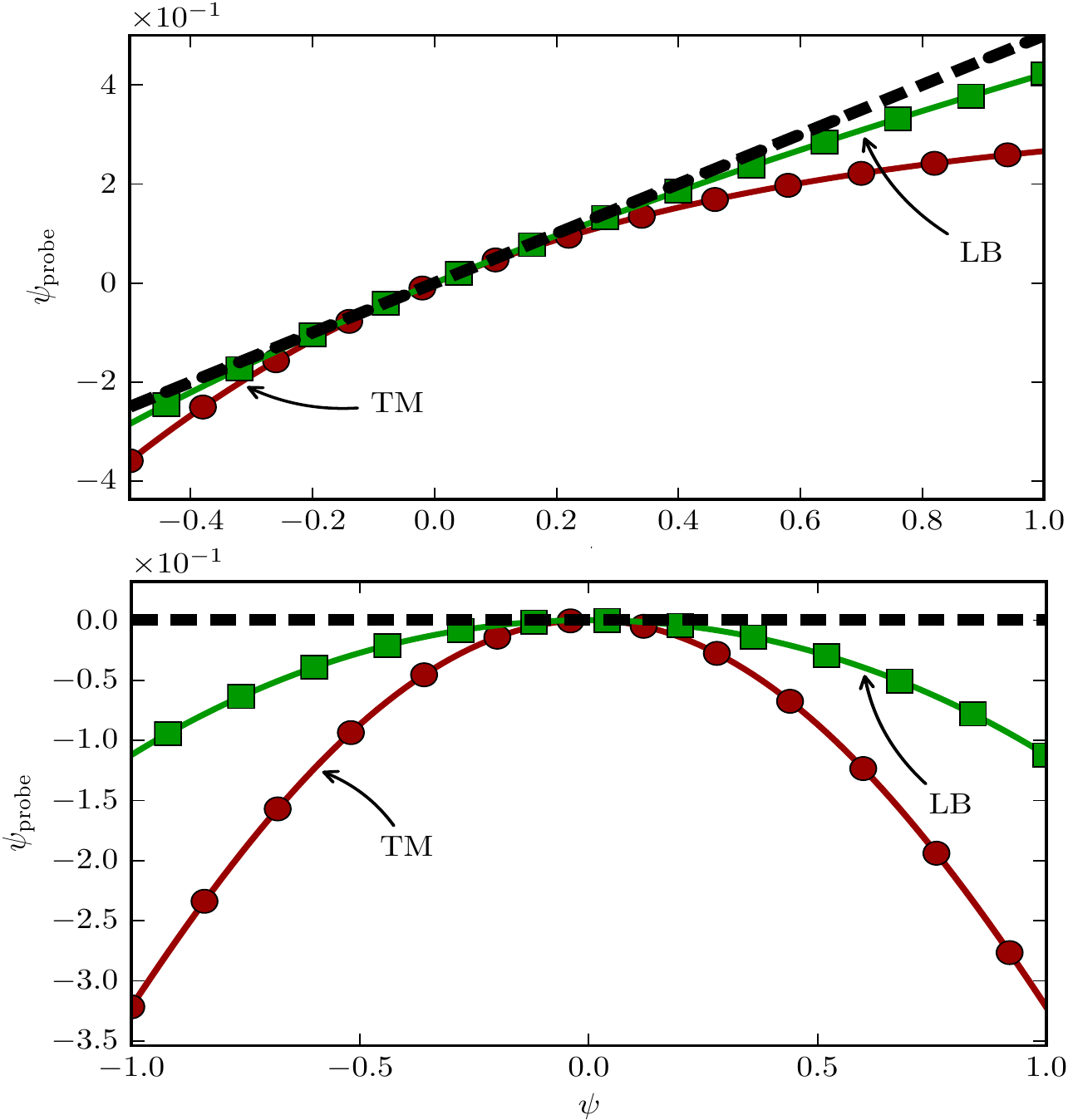}
  \caption{(Color online) Comparison of the local temperature in the TM and the LB approach. The relative local temperature differences $\psi_\probe$ are plotted against
    the relative temperature bias $\psi$. In the upper panel, the TM field or temperature bias is only applied to the left lead. The dashed, black line depicts the
    approximation $\psi_\probe \approx 0.5 \psi$, which corresponds to taking the average between the two leads as estimate for the temperature at the impurity.
    In the lower panel, the temperature bias is applied symmetrically to the left and right leads, i.e., the temperature in the right lead is change by the same amount as in the left lead
    but in the opposite direction. Accordingly, a simple estimate of the relative temperature difference at the impurity  $\psi_\probe = 0$, which is
    shown by the horizontal dashed, black line. The circles (red curve, labeled ``TM'') show $\psi_\probe$
    in the TM approach and the squares (green curve, labeled ``LB'') in the LB approach. \label{FIG:localTemperature}}
\end{figure}

In the model we have introduced in Sec.\ \ref{SEC:ThermoelectricTransport}, the tip simply corresponds to a specific
lead which we refer to as the ``probe lead'' (cf.\ Fig.\ \ref{FIG:TBModelProbe}). We assume that the decay rate of the probe lead
is much smaller than the decay rates of all the other leads, i.e.,
\begin{align}
  \frac{|V_\probe|^2}{\hbar t_\probe} \ll \frac{|V_\alpha|^2}{\hbar t_\alpha} ~, \label{probeCondition}
\end{align}
where $\alpha$ labels all other leads connected to the device. Furthermore, we take the probe lead to be
half-filled, i.e., the center of the band is aligned with the chemical potential, and that it is weakly
coupled to the impurity, i.e., $V_\probe/t_\probe \ll 1$. This means that we can treat the probe lead in the
so-called wide-band limit, which implies that the embedding self-energy due to the probe lead can be approximated by
\begin{align}
  \Sigma^{\mathrm{R/A}}_\probe(\epsilon) & = \mp \frac{i}{2} \Gamma_\probe ~, \label{probeSelfEnergyWBL}
\end{align}
i.e., it is essentially given by a characteristic frequency-independent decay rate $\Gamma_\probe / (2 \hbar)$.
Under these assumptions, the currents in the probe lead are given by
\begin{subequations} \label{probeCurrents}
  \begin{align}
    I_\probe & = \frac{1}{\hbar} \sum_{\alpha} \frac{1}{2 \pi} \indll{\epsilon}{-\infty}{\infty} \nn
    & \times \frac{\Gamma_\probe \Gamma_{\alpha}(\epsilon) \left(f_\probe - f_{\alpha} \right)}
    {\left(\epsilon - \left(\epsilon_{\mathrm{imp}} - \mu\right) - \frac{1}{2}\Lambda(\epsilon) \right)^2 + \left(\frac{1}{2}\Gamma(\epsilon)\right)^2} ~, \label{probeCurrent} \\
    Q_\probe & = \frac{1}{\hbar} \sum_{\alpha} \frac{1}{2 \pi} \indll{\epsilon}{-\infty}{\infty} \epsilon \nn
    & \times \frac{\Gamma_\probe \Gamma_{\alpha}(\epsilon) \left(f_\probe - f_{\alpha} \right)}
    {\left(\epsilon - \left(\epsilon_{\mathrm{imp}} - \mu\right) - \frac{1}{2}\Lambda(\epsilon) \right)^2 + \left(\frac{1}{2}\Gamma(\epsilon)\right)^2} ~, \label{probeHeatCurrent}
  \end{align}
\end{subequations}
where due to the assumption of a weakly coupled probe lead $\Lambda(\epsilon)$ and $\Gamma(\epsilon)$ do not include
the contribution due to the probe lead. The zero-current conditions take the neat form
\begin{subequations} \label{probeZeroCurrentConditions}
  \begin{align}
    \sum_{\alpha} \frac{1}{2 \pi} \indll{\epsilon}{-\infty}{\infty} D_\alpha(\epsilon) f_\alpha
    & = \frac{1}{2 \pi} \indll{\epsilon}{-\infty}{\infty} D(\epsilon) f_\probe ~, \label{probeZeroCurrent} \\
    \sum_{\alpha} \frac{1}{2 \pi} \indll{\epsilon}{-\infty}{\infty} \epsilon D_\alpha(\epsilon) f_\alpha
    & = \frac{1}{2 \pi} \indll{\epsilon}{-\infty}{\infty} \epsilon D(\epsilon) f_\probe ~, \label{probeZeroHeatCurrent}
  \end{align}
\end{subequations}
where we introduced the density of states
\begin{subequations} \label{DoS}
  \begin{align}
    D(\epsilon) & = \sum_\alpha D_\alpha(\epsilon) \label{densityOfStates} \\
    D_\alpha(\epsilon) & = \frac{\Gamma_{\alpha}(\epsilon)}{\left(\epsilon - \left(\epsilon_{\mathrm{imp}} - \mu\right) - \frac{1}{2}\Lambda(\epsilon) \right)^2
      + \left(\frac{1}{2}\Gamma(\epsilon)\right)^2} ~. \label{leadDensityOfStates}
  \end{align}
\end{subequations}
Note that Eqs.\ \eqref{probeZeroCurrentConditions} does not depend on the specific value of the decay rate
to the probe lead.\footnote{This is a consequence of taking the wide-band limit.} The conditions given in Eqs.\ \eqref{probeZeroCurrentConditions} determine the potential
bias $U_\probe$ and the relative temperature bias $\psi_\probe = \delta T_\probe / T$ that enter in the occupation function
$f_\probe$. It is important to realize that these equations are nonlinear, i.e., finding the potential and temperature bias
that yield vanishing particle and heat currents is nontrivial. \footnote{Equations \eqref{probeZeroCurrentConditions} have been
  derived by linearization in terms of the decay rate to the probe lead.}

Interestingly, the left-hand side of Eq.\ \eqref{probeZeroCurrent} corresponds to the long-time limit of the particle density
of the impurity site and, similarly, the left-hand side of Eq.\ \eqref{probeZeroHeatCurrent} corresponds to the long-time limit of the
energy density of the impurity site. Note that the long-time limits of the density and energy density correspond
to the long-time limit in the absence of the probe lead.
Further details on the definition of the energy density of the impurity site are given in Appendix \ref{APP:energyDensity}.
Moreover, the right-hand side of Eq.\ \eqref{probeZeroCurrent} can be interpreted as the density of the impurity site in equilibrium
with a bath at temperature $T_\probe$ and chemical potential $U_\probe$. The right-hand side of Eq.\ \eqref{probeZeroHeatCurrent}
is the energy density of the impurity site under the same equilibrium conditions. Accordingly, the zero-current conditions, Eq.\ \eqref{probeZeroCurrentConditions},
are identical to asking the question: what is the temperature and the chemical potential that reproduce the steady-state particle density
and energy density of the device under equilibrium conditions?

It should be emphasized that the potential and temperature biases applied to the other leads enter in the
definition of the density of states [cf.\ Eq.\ \eqref{densityOfStates}] via $\Gamma_\alpha(\epsilon)$ and
$\Lambda_\alpha(\epsilon)$. We remind the reader that the difference between the LB and the TM approaches manifests itself in two places. Firstly, in the TM
approach, the density of states depends on the applied TM field and potential biases, whereas in the LB approach only the applied potential biases enter in the density of states.
Secondly, the difference shows up in the way the probe potential and temperature bias appear in the occupation function, i.e.,
\begin{subequations} \label{probeOccupationFunctions}
  \begin{align}
    f_{\probe}^{\mathrm{TM}} & = f_{T}\left(\frac{\epsilon}{1 + \psi_{\probe}} - U_{\probe}\right) \nn
    & = f_{T_{\probe}}\left(\epsilon - (1 + \psi_{\probe})U_{\probe}\right) ~, \label{probeOccupationFunctionsTM} \\
    f_{\probe}^{\mathrm{LB}} & = f_{T_{\probe}}\left(\epsilon - U_{\probe}\right) ~. \label{probeOccupationFunctionsLB}
  \end{align}
\end{subequations}

As a specific example we consider the setup shown in Fig.\ \ref{FIG:TBModelProbe}. In order to simplify the analysis, we only apply a temperature
bias and employ the zero-current conditions \eqref{probeZeroCurrentConditions} to find the effective local potential $U_\probe$ and the effective
local temperature $T_\probe=T(1 + \psi_\probe)$. By symmetry, the local potential $U_\probe$ of the probe lead vanishes.
In Fig.\ \ref{FIG:localTemperature}, we show the local temperature in the TM and the LB
approaches for two scenarios. The upper panel corresponds to the situation where the temperature of the left lead is changed while the temperature of
the right lead is kept constant. The lower panel depicts a situation when the temperature in the right lead is changed by the same amount as the temperature
in the left lead, but in the opposite direction. The simplest estimate of the local potential and the local temperature is provided by the mean of the applied
potential and temperature biases, respectively; this is shown by the dashed black line in Fig.\ \ref{FIG:localTemperature}.  However, we see that in this model
and under the bias conditions described above, the local effective temperature is always lower than the average between the left and right temperature.
The deviation from the simple estimate at finite $\psi$  is much stronger in the TM approach than in  the LB approach.

The fact that the temperature of the probe is not equal to the average temperature is clearly a nonlinear effect.\cite{DubiDiVentra:09}
It is the nonuniformity of the temperature that breaks the symmetry between the left and right halves of the system and
allows the temperature of the junction to be closer to the temperature of one reservoir than to the temperature of the other.
Explicitly this can be seen from Eq.\ \eqref{probeZeroCurrentConditions}. Even if we take the density of states $D_\alpha$ to be constant, the conditions imply that we are
searching for the temperature and potential of a single Fermi function that reproduces the zeroth (density) and first moment (energy density) of a sum of Fermi functions.

\section{ Discussion and conclusion } \label{SEC:conclusion}

In this paper, we have compared the standard LB approach to thermoelectric transport to
a novel approach based on Luttinger's idea to describe temperature differences via the
TM field $\psi$. We have shown that, in fact, the TM approach encompasses the LB approach when the TM field is applied in the
initial preparation of the system. However, the TM approach allows, in addition, for a \emph{dynamical} description of
temperature variations. An interesting question is related to what extent a dynamical TM field may be realized physically.
A possible scenario in which a dynamical TM field seems to be appropriate is the process of an adiabatic compression.
In terms of the model tight-binding Hamiltonian--employed in this paper--an adiabatic compression corresponds to suddenly squeezing the atoms,
described in terms of a tight-binding chain. Since then the atoms are closer to each other the hopping
amplitude will increase, which, in turn, is precisely what the TM field, switched on at $t=0$, describes.

In the present work, we have only discussed the TM field applied to noninteracting electrons. However, we point
out that in the framework of our recently proposed thermal DFT,\cite{EichVignale:14} this suffices to address thermoelectric transport
of interacting fermions, since the interacting problem is mapped onto a noninteracting system, the so-called Kohn-Sham system.
The TM field in the Kohn-Sham system effectively describes the influence of electron-electron interactions on the
heat and particle transport. We stress that simply applying the usual time-dependent DFT\cite{RungeGross:84,GiulianiVignaleDFT:05,Ullrich:12} to thermoelectric transport
ignores the influence of the electron-electron interactions on thermoelectric transport, since by construction
it only focuses on the charge degree of freedom, i.e., the particle (charge) density. In our recently proposed thermal DFT
there will be, in general, a nontrivial TM field in the KS system  even if
there is no dynamical TM field in the physical system.

As a concrete application we have presented the computation of an effective local temperature. This is of interest for the
theoretical modeling of recent thermal scanning microscopy experiments where length scales
at which quantum mechanical oscillations become important are accessible.\cite{DubiDiVentra:09} Surely the presented results are mostly a proof of concept for
the computation of a local temperature. A more detailed investigation on more realistic devices, including multiple
states in the scattering region is currently underway. Furthermore, the employed definition of the local temperature via
a local probe under zero particle (charge) current and zero heat current seems to be somewhat artificial, since there is no
experimental ammeter for heat/energy currents (the only way to ensure the heat current vanishes is to wait for local equilibration to occur).
We are confident that the presented work paves the way for a fully microscopic
description of the combined charge and energy transport carried by electrons including the effect of electron-electron interactions via thermal DFT.

\begin{acknowledgments}
  We gratefully acknowledge support from DOE under Grant Nos. DE-FG02-05ER46203 (F. G. E., G. V.) and DE-FG02-05ER46204 (M. D.).
  F. G. E., A. P. and G. V. thank Giovanni Cicotti, Universit{\`a} di Roma La Sapienza, for the hospitality.
\end{acknowledgments}

\begin{appendix}
  
  \section{Long-time limit of the particle current} \label{APP:particleCurrent}
  
  Here, we present the explicit derivation of the long-time limit of the particle current. Combining Eqs.\ \eqref{I_G_lesser}
  and \eqref{G_lesser_0ak}, we obtain
  \begin{align}
    I_{\alpha}(t) & = \frac{\hbar}{\pi} \sum_{k} \RE\Bigg[ V_{\alpha, k} \sum_{\lambda, \lambda'}
      \indll{\epsilon}{-\infty}{\infty} f(\epsilon) \left( \bar{\mathcal{G}}^{\mathrm{A}}_{\lambda, \lambda'}(\epsilon) - \bar{\mathcal{G}}^{\mathrm{R}}_{\lambda, \lambda'}(\epsilon) \right) \nn
      & \times \frac{1}{2 \pi} \indll{\omega}{-\infty}{\infty} e^{- i  \omega(t - t_{0})} \frac{1}{2 \pi} \indll{\omega'}{-\infty}{\infty} e^{i  \omega'(t - t_{0})} \nn
      & \times \tilde{\mathcal{G}}^{\mathrm{R}}_{0, \lambda}(\hbar \omega) \tilde{\mathcal{G}}^{\mathrm{A}}_{\lambda', (\alpha, k)}(\hbar \omega') \Bigg] ~, \label{eq:app_I_alpha}
  \end{align}
  where the difference between the Green's functions $\bar{\mathcal{G}}^{\mathrm{A}/\mathrm{R}}_{\lambda, \lambda'}(\epsilon)$ and $\tilde{\mathcal{G}}^{\mathrm{A}/\mathrm{R}}_{\lambda, \lambda'}(\epsilon)$
  has been explained after Eq.\ \eqref{G_lesser_0ak}. As we have mentioned in Sec.\ \ref{SEC:SteadyState}, we will only 
  keep contributions arising from the poles of the free propagators $\bar{g}^{\mathrm{A}/\mathrm{R}}_{\alpha, k}(\epsilon)$
  and $\tilde{g}^{\mathrm{A}/\mathrm{R}}_{\alpha, k}(\epsilon)$ on which the functions $\bar{\mathcal{G}}^{\mathrm{A}/\mathrm{R}}_{\lambda, \lambda'}(\epsilon)$
  and $\tilde{\mathcal{G}}^{\mathrm{A}/\mathrm{R}}_{\lambda, \lambda'}(\epsilon)$ depend [cf.\ Eqs.\ \eqref{GF_z}]. We first consider the contribution of the
  retarded Green's function on the right-hand side of Eq.\ \eqref{eq:app_I_alpha}. Using Eq.\ \eqref{tunnelingGF_1}, we obtain
  \begin{align}
    & \lim_{t \to \infty} \frac{\hbar}{2 \pi} \indll{\omega}{-\infty}{\infty} e^{- i  \omega(t - t_{0})} \tilde{\mathcal{G}}^{\mathrm{R}}_{0, \lambda}(\hbar \omega)  \label{LTL_1} \\
    & = - i   \lim_{t \to \infty} \delta_{\lambda (\alpha', k')} V^\star_{\alpha', k'} e^{- i  \tilde\epsilon_{\alpha' , k'}(t - t_{0}) / \hbar}
    \tilde{\mathcal{G}}^{\mathrm{R}}_{0, 0}(\tilde\epsilon_{\alpha', k'}) ~, \nonumber
  \end{align}
  which tells us that the summation over ${\lambda}$ is restricted to the leads. The impurity Green's function is broadened
  by the presence of the leads [cf.\ Eq.\ \eqref{impurityGF}, Sec.\ \ref{SEC:ThermoelectricTransport}], and its contribution to Eq.\
  \eqref{LTL_1} vanishes in the long-time limit.
  Next, we consider the contribution due the advanced Green's function in Eq.\ \eqref{eq:app_I_alpha},
  \begin{align}
    & \lim_{t \to \infty} \frac{\hbar}{2 \pi} \indll{\omega}{-\infty}{\infty} e^{i  \omega(t - t_{0})} \sum_{k} V_{\alpha, k} \tilde{\mathcal{G}}^{\mathrm{A}}_{\lambda', (\alpha, k)}(\hbar \omega)  \label{LTL_2} \\
    & = i  \lim_{t \to \infty} \delta_{\lambda' (\alpha'', k'')} V_{\alpha'', k''} e^{i  \tilde\epsilon_{\alpha'' , k''}(t - t_{0}) / \hbar} \nn
    & \times \left( \delta_{\alpha \alpha''} + \tilde{\mathcal{G}}^{\mathrm{A}}_{0, 0}(\tilde\epsilon_{\alpha'', k''}) \tilde{\Sigma}^{\mathrm{A}}_{\alpha}(\tilde\epsilon_{\alpha'', k''}) \right) ~. \nonumber
  \end{align}
  To obtain this expression, we have used Eqs.\ \eqref{embeddingSE} and \eqref{leadGF}.
  The summation over $k$ allows us to identify the self-energy $\Sigma^{\mathrm{A}}_{\alpha}(\hbar \omega)$, which is a well-behaved function with no pole. Accordingly
  we can disregard the term in which $\lambda'$ refers to the impurity site in the long-time limit.
  This is the essential point in the derivation of the long-time limit as discussed in the first paragraph of Sec.\ \ref{SEC:SteadyState}.
  The interested reader may find a more careful discussion in Appendix \ref{APP:memoryLoss}.
  Using Eqs.\ \eqref{LTL_1} and \eqref{LTL_2}, we obtain the intermediate result
  \begin{align}
    & \lim_{t \to \infty} \sum_{k} V_{\alpha, k} \mathcal{G}^{<}_{0, (\alpha, k)}(t, t) \label{LTL_3} \\
    & = \lim_{t \to \infty}
    \sum_{\alpha', k'} \sum_{\alpha'', k''} V^\star_{\alpha', k'} V_{\alpha'', k''} e^{-i  \left(\tilde\epsilon_{\alpha' , k'} - \tilde\epsilon_{\alpha'' , k''}\right)(t - t_{0}) / \hbar} \nn
    & \times \frac{1}{2 \pi \hbar} \indll{\epsilon}{-\infty}{\infty} f(\epsilon) \left( \bar{\mathcal{G}}^{\mathrm{A}}_{(\alpha', k'), (\alpha'', k'')}(\epsilon)
    - \bar{\mathcal{G}}^{\mathrm{R}}_{(\alpha', k'), (\alpha'', k'')}(\epsilon) \right) \nn
    & \times \tilde{\mathcal{G}}^{\mathrm{R}}_{0, 0}(\tilde\epsilon_{\alpha', k'}) \left( \delta_{\alpha \alpha''}
    + \tilde{\mathcal{G}}^{\mathrm{A}}_{0, 0}(\tilde\epsilon_{\alpha'', k''}) \tilde{\Sigma}^{\mathrm{A}}_{\alpha}(\tilde\epsilon_{\alpha'', k''}) \right) ~. \nonumber
  \end{align}
  
  In order to proceed, we consider the difference of the equilibrium Green's functions,
  \begin{align}
    & \bar{\mathcal{G}}^{\mathrm{A}}_{(\alpha', k'), (\alpha'', k'')}(\epsilon) - \bar{\mathcal{G}}^{\mathrm{R}}_{(\alpha', k'),(\alpha'', k'')}(\epsilon) \label{initialOccupations} \\
    & = \delta_{\alpha' \alpha''} \delta_{k' k''} \left( \bar{g}^{\mathrm{A}}_{\alpha', k'}(\epsilon) - \bar{g}^{\mathrm{R}}_{\alpha', k'}(\epsilon) \right) \nn
    & + \left( \bar{g}^{\mathrm{A}}_{\alpha', k'}(\epsilon) V_{\alpha', k'} \bar{\mathcal{G}}^{\mathrm{A}}_{0,0}(\epsilon) V^\star_{\alpha'', k''} \bar{g}^{\mathrm{A}}_{\alpha'', k''}(\epsilon) \right. \nn
    & - \left. \bar{g}^{\mathrm{R}}_{\alpha', k'}(\epsilon) V_{\alpha', k'} \bar{\mathcal{G}}^{\mathrm{R}}_{0,0}(\epsilon) V^\star_{\alpha'', k''} \bar{g}^{\mathrm{R}}_{\alpha'', k''}(\epsilon) \right) ~, \nonumber
  \end{align}
  where we have used Eq.\ \eqref{leadGF}. The first two terms yield the density of states of the uncontacted leads.
  The remaining terms are due to the fact that we are working in the partition-free
  approach to transport, i.e., the leads are at all times coupled to the impurity and, hence, the density of states is broadened due to the coupling of lead $\alpha$
  to all the other leads via the impurity site. However, in the long-time limit, the contribution due to this broadening of the leads is negligible, which is shown
  explicitly in Appendix \ref{APP:memoryLoss}.
  This implies that the steady-state current is insensitive on whether we are working in the partitioned or the partition-free approach,
  even if the broadening, present in the initial state, surely affects the transient currents.
  Accordingly, Eq.\ \eqref{LTL_3} simplifies to
  \begin{align}
    & \lim_{t \to \infty} \sum_{k} V_{\alpha, k} \mathcal{G}^{<}_{0, (\alpha, k)}(t, t) \label{LTL_4} \\
    & = \tfrac{i}{\hbar} \sum_{\alpha', k'} \left| V_{\alpha', k'} \right |^2 \indll{\epsilon}{-\infty}{\infty} f(\epsilon) \delta(\epsilon - \bar\epsilon_{\alpha', k'}) \nn
    & \times \tilde{\mathcal{G}}^{\mathrm{R}}_{0, 0}(\tilde\epsilon_{\alpha', k'}) \left( \delta_{\alpha \alpha'}
    + \tilde{\mathcal{G}}^{\mathrm{A}}_{0, 0}(\tilde\epsilon_{\alpha', k'}) \tilde{\Sigma}^{\mathrm{A}}_{\alpha}(\tilde\epsilon_{\alpha', k'}) \right) ~. \nonumber
  \end{align}
  It is important to realize that the $\delta$ function due to the density of states contains the equilibrium dispersion, whereas all other Green's functions
  and self-energies contain ${\tilde{\epsilon}_{\alpha, k}}$. However, we can shift the integration variable ${\epsilon \to \frac{\epsilon}{1 + \psi_{\alpha'}} - U_{\alpha'}}$
  to transform the dispersion in the $\delta$ function into ${\tilde{\epsilon}_{\alpha', k'}}$.\footnote{Note that the change in integral weight is compensated by
    extracting a corresponding scale factor $1 +\psi_{\alpha'}$ from the $\delta$ function.} Defining
  \begin{align}
    \Gamma_{\alpha}(\epsilon) \equiv 2 \IM\left[\tilde{\Sigma}_{\alpha}^{\mathrm{A}}(\epsilon)\right]
    = 2 \pi \sum_k \left| V_{\alpha, k} \right|^2 \delta(\epsilon - \tilde{\epsilon}_{\alpha, k}) ~, \label{Gamma_def}
  \end{align}
  we arrive at
  \begin{align}
    & \lim_{t \to \infty} \sum_{k} V_{\alpha, k} \mathcal{G}^{<}_{0, (\alpha, k)}(t, t) \label{LTL_5} \\
    & = \tfrac{i}{\hbar} \sum_{\alpha'} \frac{1}{2 \pi} \indll{\epsilon}{-\infty}{\infty} f\left(\frac{\epsilon}{1+\psi_{\alpha'}}-U_{\alpha'}\right) \nn
    & \times \Gamma_{\alpha'}(\epsilon) \tilde{\mathcal{G}}^{\mathrm{R}}_{0, 0}(\epsilon) \left( \delta_{\alpha \alpha'}
    + \tilde{\mathcal{G}}^{\mathrm{A}}_{0, 0}(\epsilon) \tilde{\Sigma}^{\mathrm{A}}_{\alpha}(\epsilon) \right) ~. \nonumber
  \end{align}
  Introducing the abbreviation
  \begin{align}
    f_{\alpha} = f_T\left(\frac{\epsilon}{1+\psi_{\alpha}}-U_{\alpha}\right) ~, \label{shiftedOccupations}
  \end{align}
  for the shifted occupations, and
  \begin{align}
    \Lambda_{\alpha}(\epsilon) = 2 \RE\left[\tilde{\Sigma}_{\alpha}^{\mathrm{A/R}}(\epsilon)\right] ~, \label{Lambda_def}
  \end{align}
  for the real part of the embedding self-energies, we get the final form for the long-time limit of the particle current:
  \begin{align}
    I_{\alpha} & \equiv \lim_{t \to \infty} I_{\alpha}(t) = \frac{1}{\hbar} \sum_{\alpha'} \frac{1}{2 \pi} \indll{\epsilon}{-\infty}{\infty} f_{\alpha'} \nn
    & \times \frac{\Gamma_{\alpha'}(\epsilon)\Gamma(\epsilon) \delta_{\alpha \alpha'} - \Gamma_{\alpha'}(\epsilon)\Gamma_{\alpha}(\epsilon)}
    {\left(\epsilon - \left(\epsilon_{\mathrm{imp}} - \mu\right) - \frac{1}{2}\Lambda(\epsilon) \right)^2 + \left(\frac{1}{2}\Gamma(\epsilon)\right)^2} ~. \label{I_LTL}
  \end{align}
  
  \section{Derivation of the initial state memory loss} \label{APP:memoryLoss}
  
  In this appendix, we show explicitly that the broadening of the density of states in the initial
  equilibrium density matrix can be neglected in the long-time limit. Let us first recall that the broadening
  is due to the fact that the initial ensemble is computed in the presence of the tunneling amplitudes
  $V_{\alpha, k}$. In Appendix \ref{APP:particleCurrent} we have seen that this results in a contribution of the form
  \begin{align}
    \bar{g}^{\mathrm{A/R}}_{\alpha', k'}(\epsilon) V_{\alpha', k'} \bar{\mathcal{G}}^{\mathrm{A/R}}_{0,0}(\epsilon) V^\star_{\alpha'', k''} \bar{g}^{\mathrm{A/R}}_{\alpha'', k''}(\epsilon) ~, \label{initialBroadening}
  \end{align}
  to Eq.\ \eqref{initialOccupations}, which adds to the density of states due to the bare Green's function. We stress that in Eq.\ \eqref{initialBroadening} all the Green's functions are
  either retarded or advanced. When the contribution of Eq.\ \eqref{initialBroadening} is plugged into Eq.\ \eqref{LTL_3}, it combines with the two summation
  over $(\alpha', k')$ and $(\alpha'', k'')$.
  The resulting term can be written as
  \begin{align}
    & \sum_{\alpha', k'} \sum_{\alpha'', k''} V^\star_{\alpha', k'} V_{\alpha'', k''} e^{-i  \left(\tilde\epsilon_{\alpha' , k'} - \tilde\epsilon_{\alpha'' , k''}\right)(t - t_{0}) / \hbar} \nn
    & \times \bar{g}^{\mathrm{A/R}}_{\alpha', k'}(\epsilon) V_{\alpha', k'} \bar{\mathcal{G}}^{\mathrm{A/R}}_{0,0}(\epsilon) V^\star_{\alpha'', k''} \bar{g}^{\mathrm{A/R}}_{\alpha'', k''}(\epsilon) \nn
    & \mathcal{F}(\tilde\epsilon_{\alpha' , k'},\tilde\epsilon_{\alpha'' , k''}) ~, \label{MemoryLoss1}
  \end{align}
  where the function $\mathcal{F}(\tilde\epsilon_{\alpha' , k'},\tilde\epsilon_{\alpha'' , k''})$ is [cf.\ Eq.\ \eqref{LTL_3}]
  \begin{align}
    \mathcal{F}(\tilde\epsilon_{\alpha' , k'},\tilde\epsilon_{\alpha'' , k''}) &= \delta_{\alpha \alpha''}\tilde{\mathcal{G}}^{\mathrm{R}}_{0, 0}(\tilde\epsilon_{\alpha', k'}) \nn
    & + \tilde{\mathcal{G}}^{\mathrm{R}}_{0, 0}(\tilde\epsilon_{\alpha', k'})\tilde{\mathcal{G}}^{\mathrm{A}}_{0, 0}(\tilde\epsilon_{\alpha'', k''})
    \tilde{\Sigma}^{\mathrm{A}}_{\alpha}(\tilde\epsilon_{\alpha'', k''}) ~. \label{F}
  \end{align}
  and represents the contribution due to the Green's function of the time propagation. Next, we use the identity
  \begin{align}
    & \sum_{k} \left| V_{\alpha, k} \right|^2 \mathcal{F}({\tilde \epsilon}_{\alpha, k}) \nn
    & = \indll{\epsilon}{-\infty}{\infty} \sum_{k} \left| V_{\alpha, k} \right|^2 \delta(\epsilon - {\tilde \epsilon}_{\alpha, k})\mathcal{F}(\epsilon) \nn
    & = \frac{1}{2 \pi} \indll{\epsilon}{-\infty}{\infty} \Gamma_{\alpha}(\epsilon)\mathcal{F}(\epsilon) ~, \label{densityOfStateMethod}
  \end{align}
  twice to transform Eq.\ \eqref{MemoryLoss1} into
  \begin{align}
    & \sum_{\alpha'} \sum_{\alpha''} \frac{1}{2 \pi} \indll{\epsilon'}{-\infty}{\infty} e^{-i  \epsilon'(t - t_{0}) / \hbar}
    \frac{1}{2 \pi} \indll{\epsilon''}{-\infty}{\infty} e^{i  \epsilon''(t - t_{0}) / \hbar} \nn
    & \times \Gamma_{\alpha'}(\epsilon')  \Gamma_{\alpha''}(\epsilon'') \mathcal{F}(\epsilon',\epsilon'') \bar{\mathcal{G}}^{\mathrm{A/R}}_{0,0}(\epsilon) \nn
    & \times \bar{g}^{\mathrm{A/R}}_{\alpha', \epsilon'}(\epsilon) \bar{g}^{\mathrm{A/R}}_{\alpha'', \epsilon''}(\epsilon) ~. \label{MemoryLoss2}
  \end{align}
  Note that we have used Eq.\ \eqref{densityOfStateMethod} to replace $\bar{\epsilon}_{\alpha', k'} \to \epsilon'/(1 + \psi_{\alpha'})- U_{\alpha'}$
  and $\bar{\epsilon}_{\alpha'', k''} \to \epsilon''/(1 + \psi_{\alpha''})- U_{\alpha''}$. Furthermore, we have defined
  \begin{align}
    \bar{g}^{\mathrm{A/R}}_{\alpha', \epsilon'}(\epsilon) & = \frac{1 + \psi_{\alpha'}}{(1 + \psi_{\alpha'})(\epsilon + U_{\alpha'}) - \epsilon' \mp i  \eta} \nn
    & = -\bar{g}^{\mathrm{R/A}}_{\alpha',\epsilon+U_{\alpha'}}\left(\frac{\epsilon'}{1 + \psi_{\alpha'}}\right) ~. \label{GFkaToEpsilon}
  \end{align}
  We can see that when ${\bar{g}^{\mathrm{A/R}}_{\alpha', \epsilon'}(\epsilon)}$ is viewed as a function of ${\epsilon'}$ it changes its character from advanced to retarded and vice versa.
  The exponential functions on the right-hand side of Eq.\ \eqref{MemoryLoss2} force us to close the contour in the lower half of the complex plane for ${\epsilon'}$ and in the upper half
  for ${\epsilon''}$. Since the poles of the free propagators occur on the same side of the complex plane, the product of the two integrals
  vanishes identically in the long-time limit.
  
  \begin{figure}
    \includegraphics[width=.48\textwidth]{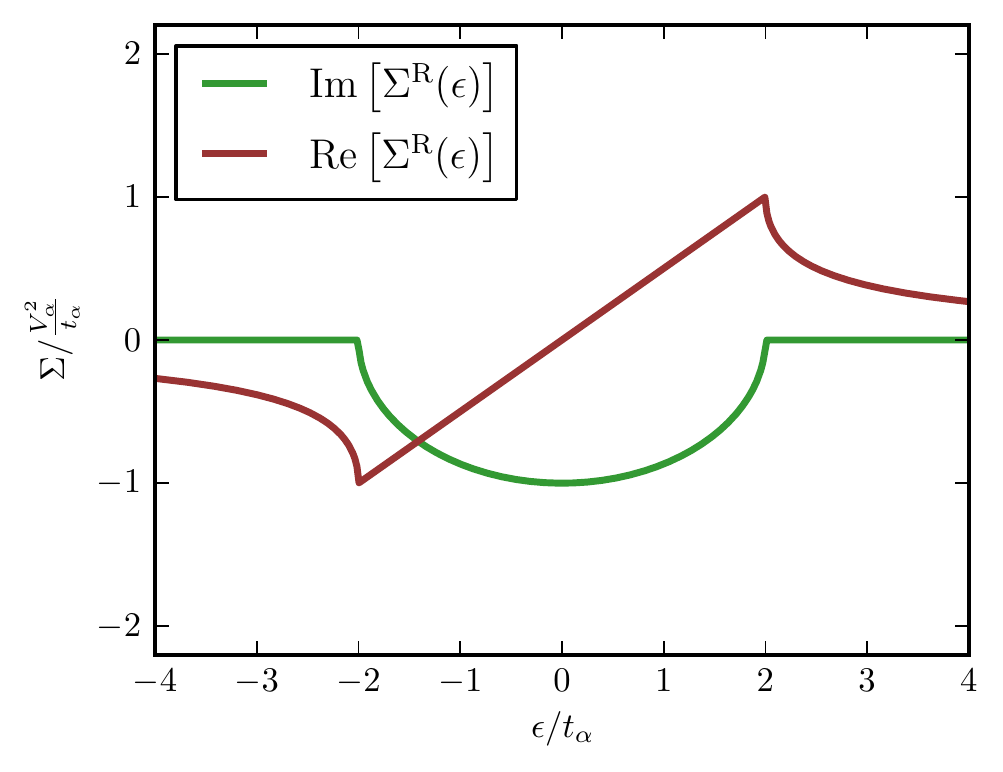}
    \caption{(Color online) ) Plot showing the real and imaginary part of the embedding self-energy for a lead modeled by a tight-binding chain
      in the limit of infinite sites. \label{FIG:SigmaChain}}
  \end{figure}
  
  We conclude this appendix by pointing out that the crucial assumption--underlying the results of the long-time limit presented in this appendix \emph{and}
  in Sec.\ \eqref{SEC:SteadyState}--is that the imaginary part of the embedding self-energy ${\Gamma_{\alpha}(\epsilon)}$ is a well-behaved function.
  Formally, however, it is given as a sum over $\delta$ functions peaked at the dispersion $\epsilon_{\alpha, k}$. It is crucial to take a continuum limit before the long-time limit.
  In the continuum limit, \emph{all} energies $\epsilon_{\alpha, k}$ get infinitesimally close, i.e., there are no bound states outside the continuum.
  As an example, we show in Fig.\ \ref{FIG:SigmaChain} the real and imaginary parts of the embedding self-energy due to a lead modeled by
  an \emph{infinite} tight-binding chain. In all numeric calculations, we have used this tight-binding model for the leads.

  \section{Long-time limit of the energy current} \label{APP:energyCurrent}

  Combining Eqs.\ \eqref{J_G_lesser} and \eqref{G_lesser_0ak}, we get the following expression for the energy current:
  \begin{align}
    J_{\alpha}(t) & = \frac{\hbar}{\pi}  \sum_{k,\lambda, \lambda'} \bar\epsilon_{\alpha, k} \RE\Bigg[ V_{\alpha, k} \nn
      & \times {} \indll{\epsilon}{-\infty}{\infty} f(\epsilon) \left( \bar{\mathcal{G}}^{\mathrm{A}}_{\lambda, \lambda'}(\epsilon) - \bar{\mathcal{G}}^{\mathrm{R}}_{\lambda, \lambda'}(\epsilon) \right) \nn
      & \times \frac{1}{2 \pi} \indll{\omega}{-\infty}{\infty} e^{- i  \omega(t - t_{0})} \frac{1}{2 \pi} \indll{\omega'}{-\infty}{\infty} e^{i  \omega'(t - t_{0})} \nn
      & \times \tilde{\mathcal{G}}^{\mathrm{R}}_{0, \lambda}(\hbar \omega) \tilde{\mathcal{G}}^{\mathrm{A}}_{\lambda', (\alpha, k)}(\hbar \omega') \Bigg] ~. \label{eq:app_J_alpha}
  \end{align}
  Formally, the only difference compared to Eq.\ \eqref{eq:app_I_alpha} is the inclusion of the equilibrium dispersion $\bar\epsilon_{\alpha, k}$ in the summation over ${k}$.
  Hence the derivation of the long-time limit of the energy current ${J_{\alpha} \equiv \lim_{t \to \infty} J_{\alpha}(t)}$ proceeds along the same lines as the calculation of
  the particle current ${I_{\alpha}}$ performed in Appendix \ref{APP:particleCurrent}.
  We emphasize that the presence of the additional factor $\bar\epsilon_{\alpha, k}$ does not alter the analytic properties determining which terms
  remain in the long-time limit. For example, Eq.\ \eqref{LTL_2} now turns into
  \begin{align}
    & \lim_{t \to \infty} \frac{\hbar}{2 \pi} \indll{\omega}{-\infty}{\infty} e^{i  \omega(t - t_{0})}
    \sum_{k} \bar{\epsilon}_{\alpha, k} V_{\alpha, k} \tilde{\mathcal{G}}^{\mathrm{A}}_{\lambda', (\alpha, k)}(\hbar \omega)
    \label{LTL_2_energy} \\
    & = i  \lim_{t \to \infty} \delta_{\lambda' (\alpha'', k'')} V_{\alpha'', k''} e^{i  \tilde\epsilon_{\alpha'' , k''}(t - t_{0}) / \hbar} \nn
    & \times \Bigg( \delta_{\alpha \alpha''} \left( \tfrac{\tilde{\epsilon}_{\alpha'', k''}}{1 + \psi_{\alpha''}} - U_{\alpha''} \right)
    + \tilde{\mathcal{G}}^{\mathrm{A}}_{0, 0}(\tilde\epsilon_{\alpha'', k''}) \nn
    & \phantom{\times} \times \left(\tfrac{1}{1 + \psi_{\alpha}} \tilde{\mathcal{T}}^{\mathrm{A}}_{\alpha}(\tilde\epsilon_{\alpha'', k''})
    - U_{\alpha} \tilde{\Sigma}^{\mathrm{A}}_{\alpha}(\tilde\epsilon_{\alpha'', k''})\right) \Bigg) ~. \nonumber
  \end{align}
  Here we have replaced the equilibrium dispersion $\bar{\epsilon}_{\alpha, k}$
  with the dispersion effective during time evolution $\tilde{\epsilon}_{\alpha, k}$ by using the relation
  \begin{align}
    \bar{\epsilon}_{\alpha, k} = \frac{\tilde{\epsilon}_{\alpha, k}}{1 + \psi_{\alpha}} - U_{\alpha} ~. \label{equilibrium_propagationRelation}
  \end{align}
  
  Compared to Eq.\ \eqref{LTL_2} the first term of Eq.\ \eqref{LTL_2_energy} has acquired an additional factor due to the equilibrium dispersion, while
  in the second term the advanced self-energy has been replace by a combination of the advance ``energy-scaled''
  self-energy ${\tilde{\mathcal{T}}^{\mathrm{A}}_{\alpha}}$ and the usual self-energy. Explicitly, we have defined
  \begin{align}
    \mathcal{T}^{\mathrm{A/R}}_{\alpha}(\epsilon) & \equiv \sum_k \epsilon_{k} \left| V_{\alpha, k} \right|^2 g^{\mathrm{A/R}}_{\alpha, k}(\epsilon) \nn
    & = \sum_{k} \left| V_{\alpha, k} \right|^2 \frac{\epsilon_{\alpha, k} - \epsilon + \epsilon}{\epsilon - \epsilon_{\alpha, k} \mp i \eta} \nn
    & = \epsilon \Sigma^{\mathrm{A/R}}_{\alpha}(\epsilon) - \sum_{k} \left| V_{\alpha, k} \right|^2 \nn
    & = \epsilon \Sigma^{\mathrm{A/R}}_{\alpha}(\epsilon) - \lim_{\epsilon \to \infty} \epsilon \Sigma^{\mathrm{A/R}}_{\alpha}(\epsilon) ~. \label{TauDefinition}
  \end{align}
  The ``tilde'' of $\tilde{\mathcal{T}}^{\mathrm{A}}_\alpha$ in Eq.\ \eqref{LTL_2_energy} indicates, by our convention, the use of the bare Green's function $\tilde{g}_{\alpha, k}(\epsilon)$
  \emph{and} the dispersion ${\tilde{\epsilon}_{\alpha, k}}$ in definition Eq.\ \eqref{TauDefinition}.
  
  The steps from Eqs.\ \eqref{LTL_3}--\eqref{LTL_5} can be directly repeated for the energy current without any additional complications. The final expression for the
  energy current only involves the imaginary part of ${\tilde{\mathcal{T}}_{\alpha}}$ and, hence, can be written in the following compelling form:
  \begin{align}
    J_{\alpha} & = \frac{1}{\hbar} \sum_{\alpha'} \frac{1}{2 \pi} \indll{\epsilon}{-\infty}{\infty} \left( \frac{\epsilon}{1+\psi_{\alpha}} - U_{\alpha} \right) f_{\alpha'} \nn
    & \times \frac{\Gamma_{\alpha'}(\epsilon)\Gamma(\epsilon) \delta_{\alpha \alpha'} - \Gamma_{\alpha'}(\epsilon)\Gamma_{\alpha}(\epsilon)}
    {\left(\epsilon - \left(\epsilon_{\mathrm{imp}} - \mu\right) - \frac{1}{2}\Lambda(\epsilon) \right)^2 + \left(\frac{1}{2}\Gamma(\epsilon)\right)^2} ~. \label{J_LTL_1}
  \end{align}
  
  \section{Definition of the impurity energy density} \label{APP:energyDensity}
  
  We define the intrinsic (kinetic) contribution to the energy density of the impurity as
  \begin{align}
    h_0(t) & = \tfrac{1}{2} \sum_{\alpha, k}  \left( V_{\alpha, k} \left\langle  \FFd{\alpha, k}(t) \FF{0}(t) \right\rangle + \mathrm{h.c.} \right) ~, \label{h0Definition}
  \end{align}
  where we adopt the convention to assign \emph{half} the energy associate with the hopping from the leads to the device
  to the impurity site. This implies that the calculation of the intrinsic impurity energy density is actually the same as
  the calculation of the particle current, i.e., we have
  \begin{align}
    \sum_{\alpha, k} V_{\alpha, k} \mathcal{G}^{<}_{0, (k, \alpha)}(t, t) & = \tfrac{1}{2} \sum_{\alpha} I_{\alpha}(t) + \tfrac{i}{\hbar} h_0(t) ~. \label{GlesserCurrentsImpurityEnergy}
  \end{align}
  From the result obtained in Appendix \ref{APP:particleCurrent}, we get the long-time limit $h_0 = \lim_{t \to \infty} h_{0}(t)$ of the intrinsic impurity energy density,
  \begin{align}
    h_0 & = \sum_{\alpha} \frac{1}{2 \pi} \indll{\epsilon}{-\infty}{\infty} f_{\alpha} \left[\epsilon - (\epsilon_{\mathrm{imp}} - \mu) \right] D_\alpha(\epsilon) ~, \label{h0_LTL} \\
    D_\alpha(\epsilon) & = \frac{\Gamma_\alpha(\epsilon)}
    {\left(\epsilon - \left(\epsilon_{\mathrm{imp}} - \mu\right) - \frac{1}{2}\Lambda(\epsilon) \right)^2 + \left(\frac{1}{2}\Gamma(\epsilon)\right)^2} ~. \label{DoS_LTL}
  \end{align}
  The first term of Eq.\ \eqref{h0_LTL} can be identified as the long-time limit of the total energy of the impurity site $h^Q_0$ and the remaining terms correspond
  to the potential energy, which is proportional to the density of the impurity site, i.e.,
  \begin{align}
    h^Q_0 & = \sum_{\alpha} \frac{1}{2 \pi} \indll{\epsilon}{-\infty}{\infty} f_{\alpha} \epsilon D_\alpha(\epsilon) \nn
    & = h_{0} + \left(\epsilon_{\mathrm{imp}} - \mu\right) n_0 ~, \label{h0Q_LTL}
  \end{align}
  where we introduced the long-time limit of the impurity density
  \begin{align}
    n_0 & = \sum_{\alpha} \frac{1}{2 \pi} \indll{\epsilon}{-\infty}{\infty} f_{\alpha} D_\alpha(\epsilon) ~. \label{n0_LTL}
  \end{align}
  The equilibrium density and total energy density of the impurity site are given by similar expressions,
  \begin{subequations} \label{equilibriumDensities}
    \begin{align}
      \left[ n_0 \right]_{\mathrm{eq}} & = \frac{1}{2 \pi} \indll{\epsilon}{-\infty}{\infty} f_{T}(\epsilon) \left[ D(\epsilon) \right]_{\mathrm{eq}} ~, \label{n0Equilibrium} \\
      \left[ h^Q_0 \right]_{\mathrm{eq}} & = \frac{1}{2 \pi} \indll{\epsilon}{-\infty}{\infty} f_{T}(\epsilon) \epsilon \left[ D(\epsilon) \right]_{\mathrm{eq}} ~, \label{h0TEquilibrium}
    \end{align}
  \end{subequations}
  where $\left[ D(\epsilon) \right]_{\mathrm{eq}}$ indicates that $\Lambda(\epsilon)$ and $\Gamma(\epsilon)$ are evaluated at vanishing potential and TM field biases.
  
\end{appendix}

\bibliography{LuttingerThermalTransport}

\end{document}